\documentclass[journal,12pt,onecolumn,draftclsnofoot]{IEEEtran}
\usepackage{cite}
\usepackage{amsmath,amssymb,amsfonts}
\usepackage{mathtools}
\usepackage{verbatimbox} 
\usepackage{moreverb}
\usepackage{algorithmic}
\usepackage{graphicx}
\usepackage{caption}
\usepackage{textcomp}
\def\BibTeX{{\rm B\kern-.05em{\sc i\kern-.025em b}\kern-.08em
    T\kern-.1667em\lower.7ex\hbox{E}\kern-.125emX}}
\usepackage{units}
\usepackage{url}
\usepackage{cite}
\usepackage{fancyvrb}
\usepackage{listings}
\usepackage{subfig}
\usepackage{multirow,tabularx}
\usepackage{filecontents}
\usepackage{xcolor}
\usepackage{enumitem}
\definecolor{darkgreen}{rgb}{0,0.75,0}

\usepackage{hyperref}
\hypersetup{
  pdfauthor = {Mohammad M. Mansour},
  pdfcreator = {Mohammad M. Mansour},
  pdfnewwindow,
  pdffitwindow=true,
  colorlinks=true,
  linkcolor=blue,       
  citecolor=darkgreen,      
  filecolor=red,        
  urlcolor=cyan         
}

\definecolor{mygreen}{rgb}{0,0.6,0}
\definecolor{mygray}{rgb}{0.5,0.5,0.5}
\definecolor{mymauve}{rgb}{0.58,0,0.82}

\lstdefinestyle{customc}{
  belowcaptionskip=1\baselineskip,
  breaklines=true,
  frame=L,
  xleftmargin=\parindent,
  language=C,
  showstringspaces=false,
  basicstyle=\footnotesize\ttfamily,
  keywordstyle=\bfseries\color{green!40!black},
  commentstyle=\itshape\color{purple!40!black},
  identifierstyle=\color{blue},
  stringstyle=\color{orange},
}

\lstdefinestyle{customasm}{
  belowcaptionskip=1\baselineskip,
  frame=L,
  xleftmargin=\parindent,
  language=[x86masm]Assembler,
  basicstyle=\footnotesize\ttfamily,
  commentstyle=\itshape\color{purple!40!black},
}

\newcommand{\nth}[1]{{#1}{\text{th}}}

\newcommand{\specialcell}[2][l]{%
  \begin{tabular}[#1]{@{}l@{}}#2\end{tabular}}

\begin{document}

\title{A Loop-Based Methodology for Reducing Computational Redundancy in Workload Sets}
\author{Elie M. Shaccour and Mohammad~M.~Mansour,~\IEEEmembership{Senior Member,~IEEE}
\thanks{
E. Shaccour and M. M. Mansour are with the Department of Electrical
and Computer Engineering, American University of Beirut, Beirut 1107 2020,
Lebanon (e-mail: mmansour@ieee.org).}
\thanks{This work was funded by Intel's Middle East Energy Efficiency Research (MER) program and the American University of Beirut (AUB) University Research Board (URB).}
}

\markboth
{Shaccour and Mansour: A Loop-Based Methodology for Reducing Computational Redundancy in Workload Sets}
{Shaccour and Mansour: A Loop-Based Methodology for Reducing Computational Redundancy in Workload Sets}

\maketitle

\begin{abstract}
The design of general purpose processors relies heavily on a workload gathering step in which representative programs are collected from various application domains. Processor performance, when running the workload set, is profiled using simulators that model the targeted processor architecture. However, simulating the entire workload set is prohibitively time-consuming, which precludes considering a large number of programs. To reduce simulation time, several techniques in the literature have exploited the internal program repetitiveness to extract and execute only representative code segments. Existing solutions are based on reducing cross-program computational redundancy or on eliminating internal-program redundancy to decrease execution time. In this work, we propose an orthogonal and complementary loop-centric methodology that targets loop-dominant programs by exploiting internal-program characteristics to reduce cross-program computational redundancy. The approach employs a newly developed framework that extracts and analyzes core loops within workloads. The collected characteristics model memory behavior, computational complexity, and data structures of a program, and are used to construct a signature vector for each program. From these vectors, cross-workload similarity metrics are extracted, which are processed by a novel heuristic to exclude similar programs and reduce redundancy within the set. Finally, a reverse engineering approach that synthesizes executable micro-benchmarks having the same instruction mix as the loops in the original workload is introduced. A tool that automates the flow steps of the proposed methodology is developed. Simulation results demonstrate that applying the proposed methodology to a set of workloads reduces the set size by half, while preserving the main characterizations of the initial workloads.

\end{abstract}

\begin{IEEEkeywords}
Loops, micro-benchmarks, performance measurement, processor characterization, similarity measures, workload characterization
\end{IEEEkeywords}

\section{Introduction}
A processor is typically optimized for its targeted application-usage domain. The design phase begins by collecting reference workloads to optimize the processor's instruction set architecture (ISA), execution cores, and cache hierarchy. For example, if a processor is targeted for multimedia usage then the set of reference workloads are chosen from a variety of media applications like video encoding, signal processing, data mining, etc. Since most domains span a wide range of applications, we end up with a large workload set. Furthermore, a processor currently at the beginning of its design cycle will not be released for at least three years. As a result, the previously chosen programs might become obsolete at release time. Thus, computer architects study the trend of the target domain to predict where it is headed next so they can choose representative programs and include them in the set. Hence the workload collection stage generates a \emph{large} workload set. To understand and optimize the processor design, the requirements of each program must be analyzed separately. To narrow down the design features, computer architects analyze and characterize the execution behavior of each benchmark. For example, identifying high data sharing patterns might lead to designing a larger shared level-2 cache for the various cores rather than a larger independent level-1 caches. Once the general processor features are decided (ISA, memory hierarchy, number of cores, etc.), a simulator is developed to decide on the final processor features, e.g., level-1 cache size of $\unit[1]{MB}$ instead of $\unit[2]{MB}$. Accordingly, the workload set needs to be simulated with each configuration to determine the best possible outcome.

\begin{figure*}[!t]
\centering
\includegraphics[scale=1]{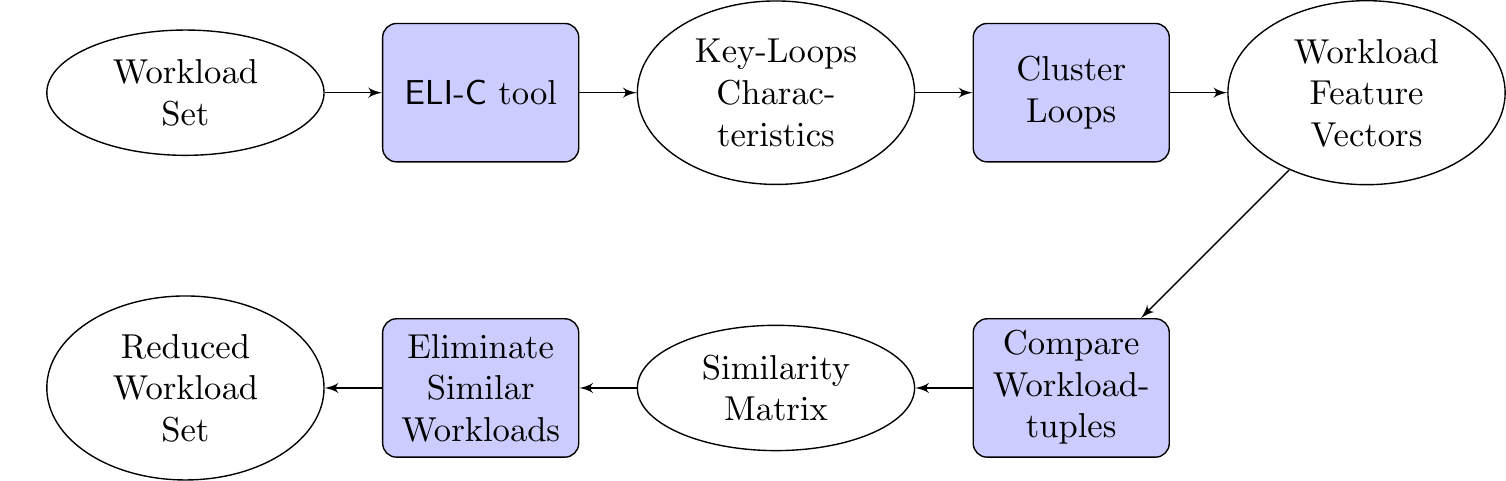}
\caption{Workflow of proposed framework} \label{fig:generalWorkFlow}
\end{figure*}

To explore further optimization opportunities, the developed simulator collects detailed simulation results from the chosen programs, and thus consumes extensive processing time and power. Some simulations require days or even months to complete, thus making it infeasible to run all the workload set with each configuration. To reduce simulation time, architects can resort to either a less detailed simulation or a reduced workload set. While less detailed simulations lead to course-grained results, an arbitrary reduction in the simulation set might mask away certain key results. Recently, researchers adopted the latter solution with systematic reduction techniques targeting the similarity among workloads. Yet, what is similarity? And how is it defined among applications?

Similarity can be measured by identifying similar patterns of operations performed by different programs. Even though applications might differ by description and by the tasks they perform, it might be evident that they interact in a similar approach with the processor such that analyzing and optimizing for only one of them would be sufficient to significantly improve the execution of the others. Being able to identify this similarity and thus remove redundant programs would reduce the workload set and thus simulation time. For example, our previous media set might include both an MPEG and a JPEG codec. Though, by description, one is for videos and the other is for images, yet the MPEG codec processes each frame using the JPEG codec, which increases redundancy in the set.

\emph{Contribution}: In this work, we propose a loop-centric workload characterization methodology together with an automation tool dubbed $\mathsf{ELI}\text{-}\mathsf{C}$ to systematically analyze and reduce a loop-dominant workload set. The proposed framework, illustrated in Fig.~\ref{fig:generalWorkFlow}, starts by identifying and characterizing key loops from the workloads in the set. Using these characterizations, we construct a feature vector for each loop and feed these vectors into a machine learning tool that classifies them into clusters based on similarity. Remapping the workloads into a vector of these cluster centers provides a systematic way to represent these programs so that they can be compared. We define a similarity score between two workloads as the inner product of the two workload vectors. Finally, we eliminate similar programs one by one using a heuristic that works with the feature scores. Using this framework, we attain a 50\% reduction in the number of programs while preserving dynamic characterization features. 

\section{Literature Review and Motivation}
Workload characterization has been employed to understand the execution behavior of multiprocessor workloads. It is well known that, due to the power-wall barrier, the trend in microprocessor development has shifted from a faster single processor core to multiple cores integrated on the same die. However, building simulators for many-core processors is a challenging task. In fact, very few architectural simulators are designed to exploit the capabilities of multiple cores offered in today's processors (e.g., MIT's Graphite simulator~\cite{miller2010graphite}). To curb simulation time and manage resource challenges for multiprocessor development projects, two complementary approaches are often adopted to reduce the target set of workloads.

The first approach is to reduce the simulation time of each application by only running smaller yet representative portions of the application. In~\cite{sherwood2001basic}, a methodology to extract a reasonably sized interval from the program that has a ``similar'' fingerprint to the overall program is presented. This is achieved by building basic block vectors (BBVs) of the program. The BBVs contain the normalized execution frequency of basic blocks (single entry, single exit portions of the code). After isolating the initialization section, clustering is performed on the remaining part of the BBVs to find the reduced set. This method exploits the internal similarity/redundancy of the programs, and aims at reducing the simulation time by using the representative program phase as a proxy for the entire program execution. Phase granularity analysis has been studied by~\cite{fang2012improving}. The SMARTS framework proposed by~\cite{wunderlich2003smarts} employs a statistical sampling strategy to predict a given cumulative property (CPI) of the entire workload with a desired confidence level. It uniformly samples the program intervals in the dynamic instruction stream for detailed simulation, and uses fast-forwarding and warm-up on the discarded instructions to speed up the simulation time. In yet another approach,~\cite{KleinOsowski_MinneSPEC_2002} minimize the input data-set of each workload in the SPEC benchmark suite to reduce simulation time.

The second approach, which relies more on characterizations, is to identify the similarity between the workloads and remove redundant programs to yield a smaller representative subset.~\cite{bienia2008parsec} presents a comprehensive comparison between PARSEC and SPLASH-2 benchmarks. Their study uses execution-driven simulations on the dynamic binary instrumentation framework (PIN) to obtain chosen characteristics including instruction mix, working set sizes, as well as the size and intensity of shared data. Using the collected information, Principle Component Analysis (PCA) is used to choose the uncorrelated data, followed by hierarchical clustering to group similar programs into single clusters. ~\cite{phansalkar2007analysis} explores the benchmark similarity inside the SPECcpu 2006 benchmark suite to identify the distinct and representative programs. They use micro-architecture dependent hardware counters to collect program characteristics. In a similar methodology to~\cite{bienia2008parsec}, they employ PCA to extract uncorrelated dimensions from a diverse set of workload characteristics. Finally, program similarity is measured using the K-means clustering approach. To retrieve workload characterizations,~\cite{bienia2008parsec} uses PIN, a detailed software simulator, while~\cite{phansalkar2007analysis} exploits the limited hardware counters approach.

Merging both previous approaches, ~\cite{eeckhout2005exploiting} presents a methodology that exploits both the program's internal repetitive phase behavior and cross-program similarities. They use SIMPOINT~\cite{perelman2003picking}, a program phase analysis tool, to break each program into intervals and collect their microarchitecture-independent characteristics. Representative phases from the entire benchmark suite are then collected. This methodology reduces the number of simulated instructions by a factor of 1.5 over SIMPOINT.

Previous approaches assume benchmarks are open source. However, most vendors are hesitant to share proprietary code, making it infeasible to use an approach that exposes details about the source code of the workload. To overcome this, a new benchmark that hides the functional meaning of the original proprietary code yet exhibits similar performance is synthesized in~\cite{joshi2008distilling,ganesan2013automatic,eeckhout2004control}. The program is first characterized, and then its characteristics are fed to a code generator that creates a program with matching characteristics.

Our proposed methodology targets loop-dominant programs. In an approach similar to~\cite{eeckhout2005exploiting}, we exploit internal as well as cross-program repetitiveness. Rather than being solely based on execution dynamics, we focus our analysis on critical program structures such as loops, to which a large percentage of a program's execution time~\cite{de2001runtime} is attributed. Internal program repetitiveness, known as the locality of reference property, is often stated in the 90/10 rule of thumb, which means that a program typically spends 90\% of its execution time in only 10\% of its static code~\cite{joshi2008distilling}. Studying loops not only provides additional views, accuracy and lower simulation time, but also opens up a new possibility to explore emerging programs using a combination of these structures. 

Isolating loops out of an entire benchmark ensures eliminating any resemblance to proprietary codes. A number of researchers have looked at loop-centric analyses. The authors in~\cite{moseley2007identifying} present a set of loop-centric profiling tools; an instrumentation-based loop profiler, which uses basic blocks to detect and account for the loops, and a light-weight sampling-based loop profiler, which has a faster profiling speed but provides less detailed information. The goal of their profiler is to help exploit loop-level parallelism and expose program hotspots. Our loop-level analysis differs from other loop-centric schemes in that it has the capability of extracting both microarchitecture-independent~\cite{hoste2007microarchitecture} as well as dependent characteristics such as arithmetic intensity, instruction mix, data reuse patterns, branch misprediction ratios, cache miss rates, etc. Once characterized, the loops then undergo PCA, which has been shown to produce the most representative subsets when more than 17 programs are used~\cite{yi2006evaluating}. This leads to better absolute and relative accuracy of CPI and energy-delay product (EDP)~\cite{yi2006evaluating}. Besides the aforementioned advantages, the objective of our framework is to allow developers to seamlessly locate and identify opportunities for execution path optimizations without much effort. Finally, we devised a generic analysis and simulation methodology based on the critical characterized loops. 

The main contributions of the paper are:
\begin{itemize}
\item Reduce profiling/characterization time of workloads
\item Propose advanced loop detection and isolation technique based on a novel metric
\item Analyze benchmarks at the binary level to reduce IP restrictions
\item Generate a micro-benchmark for each workload based on the instruction mix of the initial workload
\end{itemize} 

\section{Characterization Framework}
Traditional characterization approaches either execute every instruction of the program or sample execution periods. These approaches disregard program behavior characteristics. This simplistic approach leads to a very slow characterization process since each assembly instruction is characterized during execution. For faster approaches, some have used processor-based counters to collect characterization data. Though faster, processor counters make it difficult to establish a clear mapping between the assembly code and their characterizations. The choice between the mentioned approaches depends on whether one needs more details in a slower solution or faster, less detailed characterization. In this work, we first study the nature of the targeted applications. Using these findings we then design a framework that can efficiently provide more details at lower computational cost.

\begin{figure}[hbtp]
	\centering
		\includegraphics[scale=0.55]{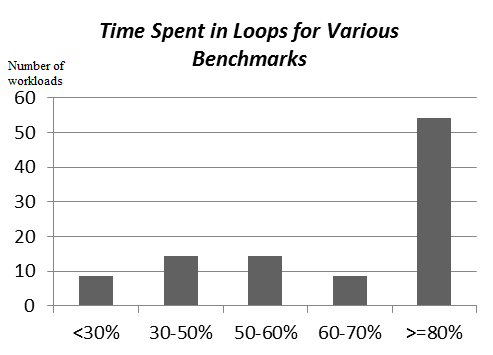}\vspace{-0.05in}
	\caption{Relative time spent in loops by workloads.}
	\label{fig:workloadsLoopTime}
\end{figure}

To understand program behavior, 108 applications are studied from many open source workload sets, including, but not limited to the following: NPB~\cite{barszcz1991parallel}, PARSEC~\cite{bienia2009parsec}, and HPCC~\cite{luszczek2006hpc}. The goal of this study is to understand the dominant programming structures in workloads to help optimize the characterization process. The relative time spent in different programming structures is recorded as well. As shown in Fig.~\ref{fig:workloadsLoopTime}, more than 50\% of programs spend more than 80\% of their time executing loops. As a result, the first hypothesis is drawn: \textit{characterizing loops alone might be sufficient to understand the entire workload, in loop-dominant applications}.

\begin{figure}[t]
	\centering
	\begin{tabular}{c}
\begin{lstlisting}[lastline=2]
for i in range(0, len(array)-1):
      sum += array[i]
\end{lstlisting}
	\end{tabular}
	\caption{Pseudocode of a linear loop}
		\label{code:linearLoop}	
		\vspace{-2em}
\end{figure}

\begin{figure}[t]
\centering
  \mbox{\subfloat[Initial nested loop]{\label{fig:nestedInitial}
\begin{lstlisting}[lastline=4]
 for i in range(0, len(array)-1):
     rowSum += array[i][0]
     for j in range(0, len(array)-1):
          sum += array[i][j]
\end{lstlisting}}}
  \mbox{\subfloat[Nested loops seperated]{\label{fig:nestedAnalyse}
\begin{lstlisting}[firstline=5]

 for i in range(0, len(array)-1):
     rowSum += array[i][0]

 for i, j in range(0, len(array)-1):
          sum += array[i][j]
\end{lstlisting}}}
  \caption{Pseudocode of a nested loop}
\end{figure}


\begin{figure}[t]
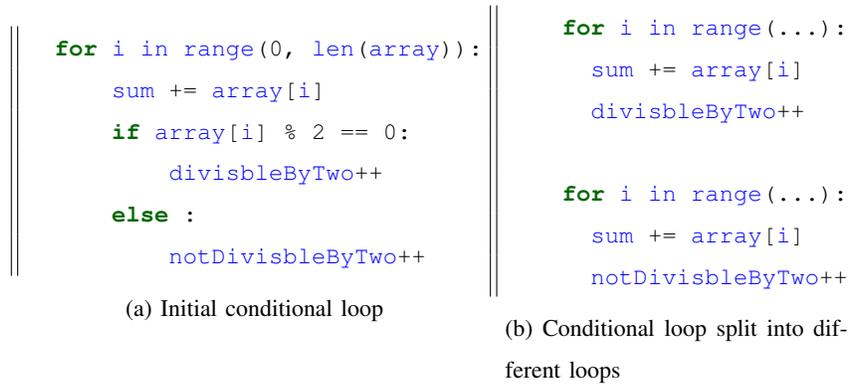

\centering
  \mbox{\subfloat[Initial conditional loop]{\label{fig:conditionalInitial}
\begin{lstlisting}[lastline=6]
  for i in range(0, len(array)):
      sum += array[i]
      if array[i] % 2 == 0:
          divisbleByTwo++
      else :
          notDivisbleByTwo++
\end{lstlisting}}}
  \mbox{  \subfloat[Conditional loop split into different loops]{\label{fig:conditionalAnalyse}
\begin{lstlisting}[firstline=8]
    for i in range(...):
      sum += array[i]
      divisbleByTwo++

    for i in range(...):
      sum += array[i]
      notDivisbleByTwo++
\end{lstlisting}}}
  \caption{Pseudocode of a conditional loop}
\end{figure}


In their simplest form, loops reduce code size by executing one or more instructions up to a desired number of times. The example shown in Fig.~\ref{code:linearLoop} is for a linear loop that calculates the sum of the elements in an array. The loop body, \texttt{sum += array[i]}, will be executed the same way as the size of the array. If the size of the array is 10, then \texttt{sum + =array[i]} will be executed 10 times. Based on the previous fact that most programs spend a large amount of time in loops and the fact that loop bodies are repetitive, the second hypothesis is drawn: \emph{characterizing a single iteration of a linear-loop and modeling it across the total number of iterations is sufficient to reduce characterization time while providing an accurate model of the entire execution}. However, what about non-linear loops?

\begin{enumerate}[itemsep=-0.2ex,leftmargin=0.5cm]
	\item \textbf{Nested-loops:} In the example shown in Fig.~\ref{fig:nestedInitial}, the outer loop accumulates the sum of the first elements in each row of the matrix, while the inner loop accumulates the sum of all the elements in the matrix. Our initial assumption fails in this case as the outer loop's body is not linear. Viewed differently, nested loops are a combination of linear loops. Identifying each loop's execution load would suffice to split a nested loop into multiple linear loops as shown in Fig.~\ref{fig:nestedAnalyse}.
	\item \textbf{Conditional statements}: In the example of Fig.~\ref{fig:conditionalInitial}, the conditional statement counts the number of elements in the array that are divisible by 2. The first statement, \texttt{sum += array[i]}, executes in each iteration, while the rest of the code is divided into separate paths and thus the code does not execute linearly. Viewed differently, each branch of the conditional statement could be considered as a separate loop. Adding to it the always-executing statement, two linear loops are formed as shown in Fig.~\ref{fig:conditionalAnalyse}.
\end{enumerate}

Performing loop splittings on the source code is a difficult task however. To overcome this, the characterization framework will analyze binary code, relying on basic block execution counts to group loops together. This has a two-folded advantage, since characterization at the binary level will give a clear view of what executes on the processor after compiler optimizations have been applied.
\begin{figure}[hbtp]
	\centering
		\includegraphics[scale=0.45]{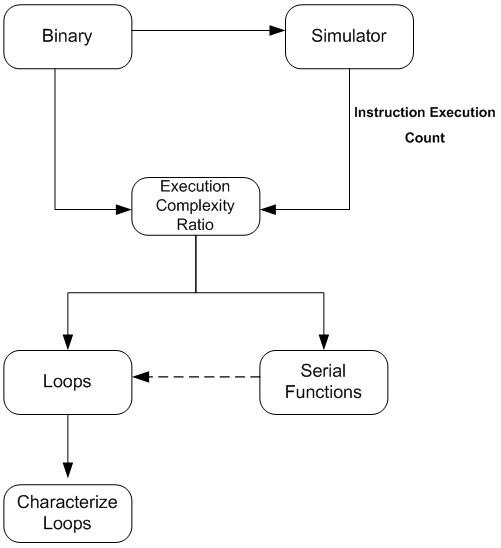}
	\caption{$\mathsf{ELI}\text{-}\mathsf{C}$ tool dataflow}
	\label{fig:toolDataFlow}
\end{figure}

\subsection{Framework data flow}
The dataflow of the proposed characterization tool ``Efficient Loop Identifier and Characterizer'' ($\mathsf{ELI}\text{-}\mathsf{C}$) is shown in Fig.~\ref{fig:toolDataFlow}. The target of the framework is to identify linear loops within a program and then characterize them. During the simulation phase, the program will be executed simultaneously using both \textsf{Gprof} \cite{graham1982gprof} and \textsf{Valgrind}~\cite{seward2004valgrind} tools. \textsf{Gprof} is used to calculate the absolute execution time spent in each function, while \textsf{Valgrind}'s internal tools, \textsf{Callgrind} and \textsf{Cachegrind}, are used to extract instruction execution counts as well as cache misses for various cache sizes, respectively. Using the instruction execution count, linear loops are identified using a new proposed ratio, dubbed Execution Complexity Ratio $ECR$, and defined as:
\begin{equation}
    ECR =\frac{AEC}{TEC}:
    \begin{cases}
        \textless 1, & \text{for conditional statement}\\
        =1, & \text{for serial code}\\
        \textgreater 1, & \text{for loop code}
\end{cases}
\label{eq:ECR}
\end{equation}
where \textit{AEC} is the Actual Execution Count or number of times the given assembly line is executed, while the \textit{TEC} is the Theoretical Execution Count or number of times the assembly line would have executed if it was part of a serial function code. Alternatively, \textit{TEC} is the number of times the function in which the instruction resides has been called. In the function shown in Fig.~\ref{fig:linearLoopNew}, each assembly instruction will execute a number of times equal to the number of times the function is called. On the other hand, in Fig.~\ref{fig:nonlinearLoopNew}, the instruction \texttt{sum += array[i]} will be executed as many times as the loop is executed multiplied by the number of times the function is called. The \textit{ECR} ratio is calculated for each assembly line. Instructions having the same \textit{ECR} and located next to each other are considered part of the same loop. Loop bodies are then isolated and characterized as a single entity. The characterizations are calculated for each loop body separately as opposed to the entire program.

\begin{figure}[t]
\centering
  \mbox{\subfloat[Linear code]{\label{fig:linearLoopNew}
\begin{lstlisting}[lastline=4]
  def FahrenheitToCelsius(f):
      temp = f - 32
      celsius = temp/1.8
      return celsius
\end{lstlisting}}}
  \mbox{\subfloat[Non-linear code]{\label{fig:nonlinearLoopNew}
\begin{lstlisting}[firstline=5]
     def SumArray(array):
      for i in range(0,len(array)):
          sum += array[i]
\end{lstlisting}}}
  \label{fig:loopsExampleNew}
  \caption{ECR detection examples}
\end{figure}

\subsection{Direct Characterizations}
Direct characterizations are characterizations computed cumulatively across each instruction in a loop, independently from other instructions. They are further divided between computational instructions and memory instructions.

\subsubsection{Characterizations of Computational Instructions}
The following parameters characterize computational instructions:
\begin{itemize}[itemsep=-0.1ex,leftmargin=0.5cm]
\item \textbf{Loop iterations}: Number of times the loop body is executed.
\item \textbf{Static instructions}: Number of assembly instructions executed in the loop per iteration.
\item \textbf{Relative size}: Static instructions are not indicative of the weight of the loop since a single-instruction loop executing a million times could be performing more computations than a 1000-instruction, 10-iteration loop. This characterization shows the relative size of the dynamic number of instructions executed by the loop relative to the total number of instructions executed by the program:
    \begin{equation*}
        \small
      \text{Relative size} = \frac{\text{Loop Iterations}\times\text{Static Instructions}}{\text{Number of Instructions Executed by Workload}}
    \end{equation*}
\item \textbf{Scalar instructions}: Number of single-operation, single-data instructions.
\item \textbf{Vector instructions}: Number of single-operation, multiple-data (SIMD) instructions.
\item \textbf{Floating-point instructions}: High latency floating-point instructions require more cycles to complete compared to integer instructions. It is essential to keep track of these instructions to understand their effect on loops.
\end{itemize}

\subsubsection{Characterizations of Memory Instructions} The following parameters characterize memory instructions:
\begin{itemize}

\item \textbf{Raw Bytes}: The previous characterizations deal with the computational aspect of loops. Memory is the other crucial part to characterize. The virtual memory system (VMS) starts with cache and goes up to the disk storage. The Raw Bytes characterization represents the sum of all data requests invoked by the loop to the VMS. It is further divided between \textbf{Bytes-Loaded} and \textbf{Bytes-Stored}.

\item \textbf{Memory Accesses}: Raw Bytes represent the data requests in bytes. Absolute numbers such as 128 bytes, might not be indicative of the processor-memory interaction since a single instruction might load the 128 bytes or it might be a one-byte load instruction executed 128 times. For a better understanding of how the memory bandwidth is used, memory accesses are separated between data loads and stores. This characterization is further divided into \textbf{Scalar Loads}, \textbf{Scalar Stores}, \textbf{Vector Loads} and \textbf{Vector Stores}.

\item \textbf{Filtered Bytes}: Another problem with the Raw Bytes characterization is that it is not representative of the behavior of the loop with respect to cache. Filtering memory references through cache gives a better understanding of data reuse within a loop. The tool is able to perform simulations for various cache sizes. This characterization is also indicative of the size of the working set.

\item \textbf{Branch Mis-prediction Rate}: To differentiate between linear and non-linear loops, one can rely on branch mis-predictions. Linear loops exhibit the lowest mis-prediction rates, followed by nested loops and conditional loops. Referring back to Fig.~\ref{fig:nestedInitial}, the loop-back instruction of the inner loop will incur higher mis-predicted branches than that of the outer loop. Given that each loop execution might have a different control flow graph, this characterization allows us to estimate the confidence with which this loop is executed.
\end{itemize}

\subsection{Indirect Characterizations}
A program's behavior cannot be captured solely using static characterisations. Indirect characterizations compensate for the limited scope of the direct characterizations by modeling the dynamic behavior of a program. Such characterizations require an extended view of previous and subsequent instructions to help classify the current instruction or the entire loop. Indirect characterizations help in understanding the intended behavior of the loop at the source code level by showing a broader view.

\begin{figure}[hbtp]
\centering
\subfloat[Stencil]{\includegraphics[width=1.0in]{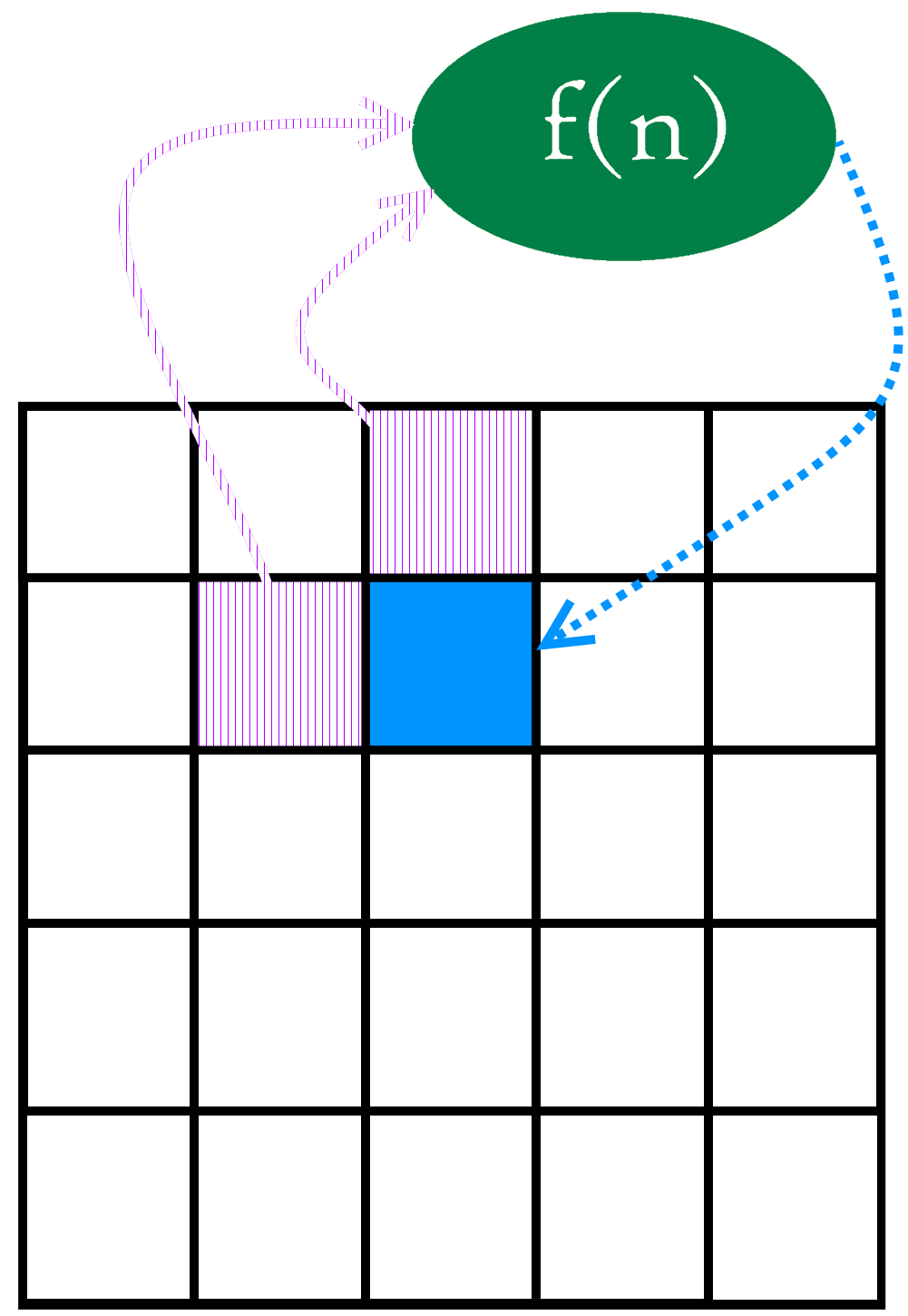}%
\label{subFig:stencil}}\\
\subfloat[Matrix transpose]{\includegraphics[width=2.0in]{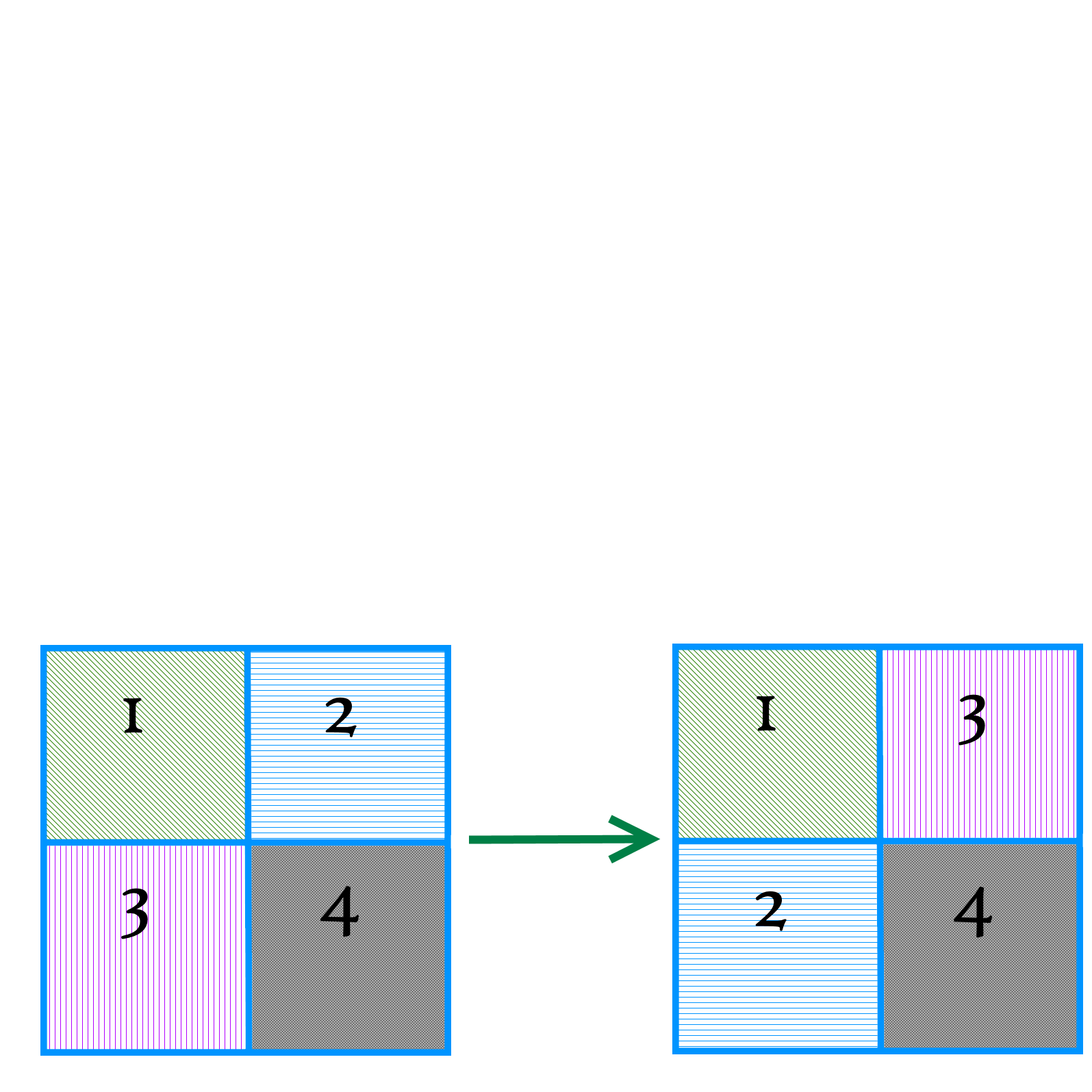}%
\label{subFig:transpose}}\\
\subfloat[Histogram]{\includegraphics[width=2.0in]{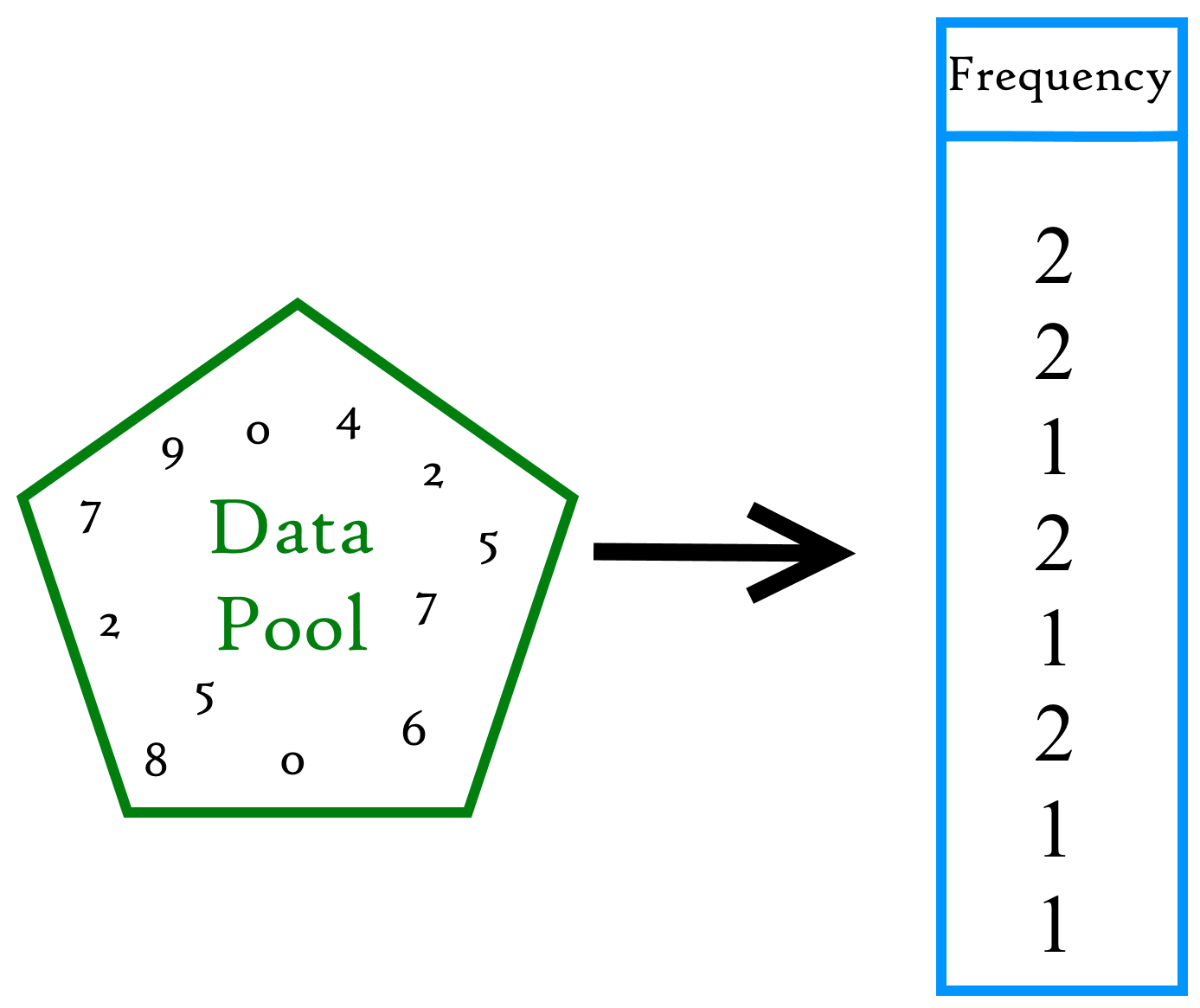}%
\label{subFig:histogram}}\\
\subfloat[Data streaming]{\includegraphics[width=2.0in]{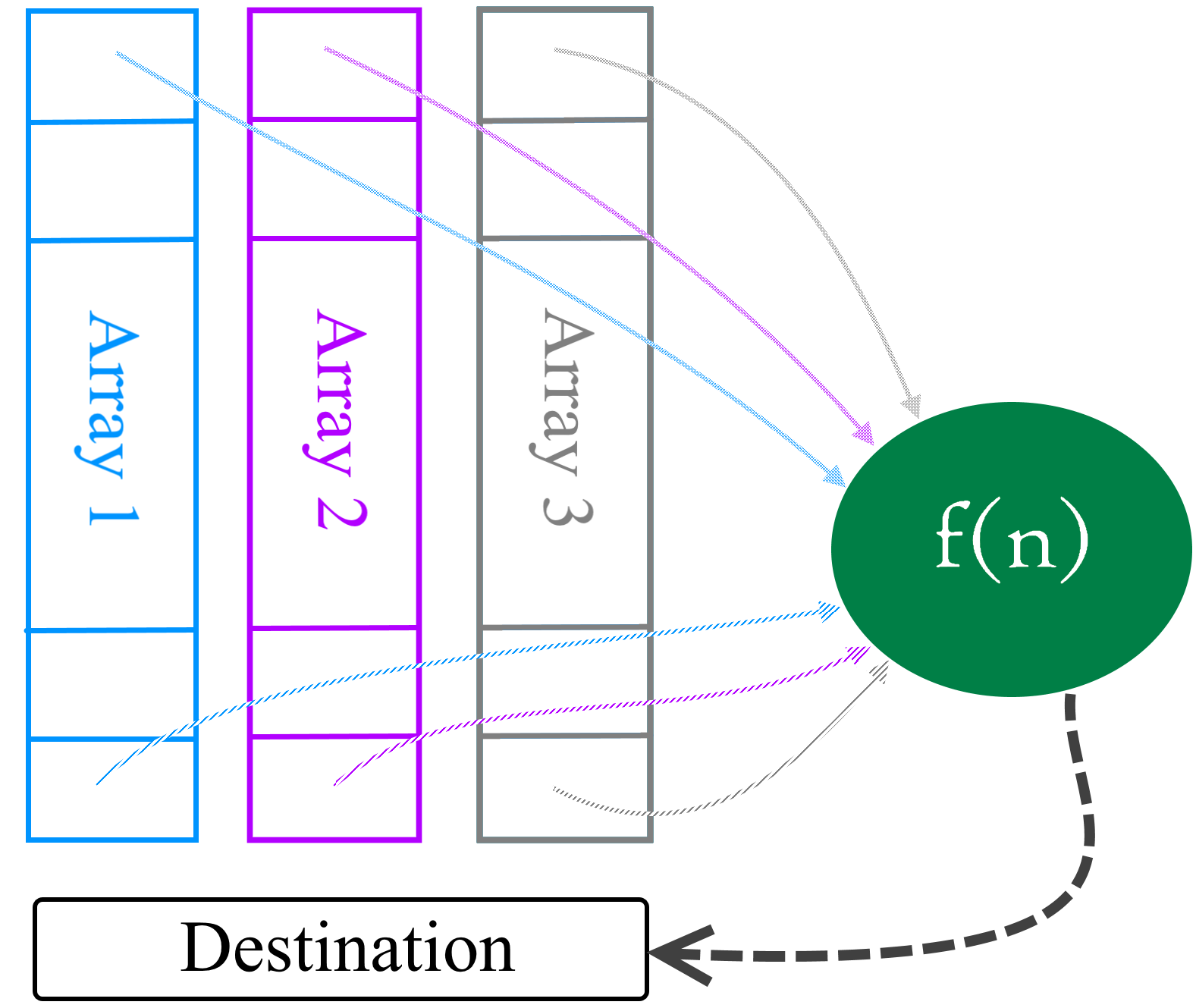}%
\label{subFig:stream}}
\caption{Computational models detected by framework: (a) Stencil, (b) matrix transpose, (c) histogram, and (d) data streaming}
\label{fig:Computational_models}
\end{figure}

\begin{itemize}[itemsep=0.25ex,leftmargin=0.5cm]

\item \textbf{Instruction classes}: Classifying instructions according to their type requires tracking the registers and their ultimate use. The $\mathsf{ELI}\text{-}\mathsf{C}$ tool creates a buffer for each loop and keeps track of registers and the way they are used. A careful breakdown of each instruction is taken into consideration since some instructions do not have a destination register (e.g. \texttt{j}), while others have the first operand also as a destination (\texttt{add}), and the rest have a separate register for destination (\texttt{mov}). The following is a description of the classes of instructions:

\begin{enumerate}[itemsep=0.25ex,leftmargin=0.5cm]

\item \emph{Address Calculation Instructions}: Architectural register limitations presented by the x86/64 architecture for example are compensated for by frequent spills to memory, and by providing more physical registers. The x86/64 ISA provides complex memory addressing modes and thus some instructions, considered as overhead instructions, are used to calculate memory addresses~\cite{intel64and}.

\item \emph{Control Instructions}: They characterize the number of instructions that change the execution flow of a loop, such as jumps and functions calls. This helps reveal, among other details, if a loop for example is part of another nested loop.

\item \emph{Compute Instructions}: Upon removing previous overhead instructions, pure compute instructions will remain. These instructions will carry out the computations required by the loop. This characterization gives an indication of the amount of computations performed by the loop.
\end{enumerate}

\item \textbf{Memory Access Patterns}: The main factor that affects the cache hit rate (Filtered Bytes), is the memory access patterns. For example, using a simple hardware cache predictor, cache misses will decrease if memory accesses of a loop follow a fixed-strided pattern. To discover memory access patterns, the tool keeps track of static addresses of both memory reads and writes, separately. For example, address $[\texttt{esp} - 5]$ is stored as is rather than replacing $\texttt{esp}$ by its actual value. Although this hides out some indirect patterns, yet it reduces analysis time. Our pattern detection algorithm is then run over the archived load and store addresses. The algorithm will detect all patterns repeating more than 2 occurrences, e.g., $-5,-5,-5$, or $-5, +6, -5, +6, -5, +6$, etc.
\item \textbf{Data Dependency}: Execution bottlenecks are usually caused by long data-dependent computations. The longest dependency chain is determined by calculating the destination register with the maximum time needed by its operands to be available in addition to the number of clock cycles (CC) required to execute the instruction~\cite{granlund2005instruction}. For example, consider the following pseudo-instruction:  \texttt{add dest, reg1, reg2}. Then,
    \begin{equation}
    t_{\texttt{dest}} = \max(t_{\texttt{reg1}}, t_{\texttt{reg2}}) + t_{\texttt{add}}
    \end{equation}
is the dependency chain from the source registers of \texttt{add} to its destination register. Once the loop executes, the longest dependency chain is calculated by traversing all registers:
\begin{equation}
\text{Longest Dependency Chain} = \max\limits_{0\leq i\leq \text{NumReg}}  t_{\texttt{dest}_{i}}.\label{eq:logn_dep_chain}
\end{equation}
Using~\eqref{eq:logn_dep_chain}, the average Instruction-Level Parallelism (ILP) within a loop can be approximated as
\begin{equation*}
  \text{ILP} \thickapprox \dfrac{\text{Number of Static Instructions}}{\text{Longest Dependency Chain}}.
\end{equation*}

\item \textbf{Computation Models}: The main disadvantage of binary characterization is that it hides the computations performed at the source code level. To better understand what the application is performing, this characterization captures and characterizes some known computational modules as defined by~\cite{asanovic2006landscape}. Loops with similar computation models will have higher similarity. Of these computations, we consider:
\begin{itemize}
\item \emph{Stencil}: Given a matrix, the current cell value is determined as a function of neighboring cells (see Fig.~\ref{subFig:stencil}).
\item \emph{Transpose}: A mathematical process whereby columns are switched to rows and vice versa (see Fig.~\ref{subFig:transpose}).
\item \emph{Histogram}: Traverse a data set counting the occurrence of unique elements (see Fig.~\ref{subFig:histogram}). Histogram data is read from memory in a unified way, while storing is done randomly.
\item \emph{Data Streaming}: Load and manipulate data from several data sets, then store results back (see Fig.~\ref{subFig:stream}). Loaded elements often show a memory access pattern that decreases cache misses.
\end{itemize}

\end{itemize}

In summary, workloads are divided into loops, which are then characterized as separate entities. The loop-entity characterization are used as a behavioral model for loop dominant workloads. Applying machine learning on these generated models will show and eliminate workload similarity as presented in the following sections. 

\section{Elimination Scheme}
Starting with a set of workloads, our target is to obtain a reduced set while maintaining the characterizations of the initial set.  As an overview, this is achieved by:
\begin{enumerate}[itemsep=-0.0ex,leftmargin=0.75cm]

\item Profiling the workloads using $\mathsf{ELI}\text{-}\mathsf{C}$

\item Creating a representative vector for each loop

\item Joining loop characterizations from the same workload to obtain a workload-specific signature vector

\item Calculating pair-wise workload similarity

\item Applying an elimination procedure on the workload set

\end{enumerate}
Each of the above steps is refined to remove biasing and improve the resulting outcome as discussed below.

\subsection{Profiling and Data Collection}
The first and simplest step is profiling the workloads using $\mathsf{ELI}\text{-}\mathsf{C}$. Assuming the workload set has $k$ loops in total, each having $n$ characterizations, the output of the profiling step will be a $k\times n$ matrix. The generated matrix is then forwarded for data preprocessing.

\subsection{Data Preprocessing}
To avoid false data analysis, as in any data mining problem, the given matrix is preprocessed to remove noise and outlier samples. Preprocessing is carried out in two complementary ways. Given the $k\times n$ matrix, the number of rows is reduced by \emph{data cleaning}, and the number of columns is reduced by \emph{data dimension reduction}.

\subsubsection{Data Cleaning}
Data cleaning removes all records that bias the analysis. This process is divided into 2 phases. The first phase removes loops that belong to non loop-dominant workloads. The second phase removes loops that exhibit low execution time relative to their workloads. Ultimately, we end up with a $(k-x)\times n$ matrix containing core loops with their characterizations, where $x$ is the total number of non-dominant loops.

\subsubsection{Data Dimension Reduction}
The gathered characterizations are split between architecture dependent and independent. Correlation between characterizations results in data biasing and ultimately leads to a biased classification. It is difficult to detect and remove correlations manually. Principle component analysis, PCA, is a commonly applied algorithm that removes correlation and reduces the dimensionality of a data set. Another source of data biasing is having different ranges for various characterizations. For example, cache miss rates vary between 0 and 1, while the number of floating-point instructions varies widely. To overcome data biasing, the collected characterizations are normalized. The normalized output is a matrix whose numerical values range between 0 and 1. The input to the PCA algorithm is the normalized matrix whose rows are the set of reduced loops, and whose columns are the collected and normalized characterizations. PCA computes new variables, called principle components (PCs), which are linear combinations of the initial characterizations. The new characterizations (i.e., the PCs) are uncorrelated, independent, and can be used to represent the loops in the classification algorithm. Yet again, the new PCs have different scalings, so they are normalized before data classification. The output of PCA will be a $(k-x)\times (n-y)$ normalized matrix, where $y$ is the difference between the number of characterizations and the PCs.

\subsection{Data Classification}
The normalized PC matrix contains representative features of each characterized loop. To classify similar loops together, an Expectation Maximization (EM) algorithm from the popular WEKA data mining toolkit~\cite{hall2009weka} is used. The popular K-means algorithm is a specialized form of an EM algorithm. EM is chosen because, unlike K-means, it decides on the optimal number of clusters. EM starts with an initial set of parameters and iterates until clustering cannot be improved any further (that is, until clustering converges or the change is sufficiently small). The \emph{expectation} step assigns objects to clusters according to the current fuzzy clustering or parameters of probabilistic clusters. The \emph{maximization} step sets the new clustering via parameters that maximize the sum of squared error (SSE) in fuzzy clustering or the expected likelihood in probabilistic model-based clustering. Once the algorithm is applied, each loop will have an additional characterization specifying the cluster it belongs to, such that the output will be a $(k-x)\times (n-y+1)$ matrix.

\subsection{Elimination}
Given the initial workload set, elimination will remove one workload from workload-tuples that are highly similar. The steps so far have concentrated around loop-level characterizations. To quantify workload-tuple similarity, a higher level of abstraction needs to be captured from the set of clustered loops. Each workload contains multiple loops, and each loop belongs to a cluster. Generating a workload-specific feature vector (FV) is achieved by iterating over the output matrix and calculating the time each workload spends in each cluster. For $n$ clusters, the FV is of size $n$, and in which each element is the normalized sum of the time spent within each cluster:
\begin{equation*}
    FV_{\text{Workload}_i \text{@} \text{Cluster}_n}= \text{Workload}_i \cap \text{Cluster}_n.
\end{equation*}
Since workloads have different execution times, each FV is then normalized over the entire execution time. From this point onward, the FVs are assumed to represent the benchmarks. Using the FVs, workload-tuple similarity is calculated and elimination proceeds. The following subsections introduce the elimination scheme step by step.

\subsubsection{Similarity Score}
Similar workloads are those that perform similar work. Quantifying the similarity between workloads is done using a similarity metric, which we define as the inner product of the feature vectors representing the two workloads. The similarity score ranges between 0 and 1, and is mathematically calculated as follow:
\begin{equation*}
\textsf{Similarity}\left(FV_m, FV_n\right)=\sum_{i=1}^{\text{ClusterSize}} FV_{m}[i]\times FV_{n}[i].
\end{equation*}

\subsubsection{Similarity Matrix}
To visualize workload similarity, a 2D $N\times N$ symmetrical matrix containing all the workloads is constructed, such that elements $(i,j)$ and $(i,j)$ in the matrix have the same value. The similarity matrix is used to choose the workload-tuples that have the highest similarity value, one of which will be eliminated.

\subsubsection{Elimination Strategy}
Workload elimination will remove one of the two workloads that are highly similar. This process is executed on the similarity matrix, such that in each iteration, the workload-tuple with the highest similarity score is chosen. Both of the chosen workloads perform similar work, one of which will be eliminated. Similarity of each workload against the remaining subset is calculated. Finally, the workload with lower similarity is removed.

\subsubsection{Elimination Limit}
Loops in the same cluster are similar and fall within ``short distance'' of each other. To maintain the dynamic characterizations of a workload set, one needs to maintain a certain number of loops within each cluster. Random elimination might lead to the removal of all loops in a cluster. The structured elimination technique monitors the number of loops within each cluster throughout the elimination iterations. Once an elimination round causes any of the cluster sizes to drop below a certain threshold (say 50\%), the eliminated workload is re-inserted and elimination terminates. The threshold value of 50\% was chosen to maintain a balance between the remaining workloads and the characteristics of the workload set.

\subsection{Validation}
To validate that the chosen subset of workloads is representative of the initial set and maintains its behavior, three dynamic execution aspects are performed between the initial and final set on different processor technologies. The validation metrics used are cycles-per-instruction (CPI), scalability, and data sharing. These metrics are collected using Intel's Sampling Enabling Product (SEP) tool version 3.5, from Intel VTune Amplifier XE 2011. 

\section{Simulation and Evaluation}
\subsection{Experimental setup}
The selected programs are compiled using Intel's C++ Compiler from the Composer XE suite 2011~\cite{iccCompiler}, then executed on $\mathsf{ELI}\text{-}\mathsf{C}$ using a system with Intel Xeon E7-4860 processors. To ensure fair workload-to-workload comparison, all programs are compiled using identical configuration flags for speed and vectorization, as well as enabling aggressive loop transformations such as fusion, block-unroll-and-jam, as well as collapsing \texttt{if statements}. Workload specific feature vectors are then processed by the WEKA tool~\cite{hall2009weka} using the EM clustering algorithm. To validate that the new workload subset is representative of the initial set, three different generations of Intel processors are used:
\begin{enumerate}[itemsep=-0.0ex,leftmargin=0.5cm]
\item \textbf{Intel Core i7-2600K}: A 64-bit, 4-core, 8-thread processor running at $\unit[3.40]{GHz}$. It supports Turbo boost 2.0 technology that can boost the clock speed up to $\unit[3.80]{GHz}$. It includes an $\unit[8]{MB}$ smart cache system and a $\unit[5]{GT/s}$ system bus.
\item \textbf{Intel Core i7-975}: A 64-bit, 4-core, 8-thread processor running at $\unit[3.33]{GHz}$. It supports Turbo boost technology that can boost clock speed up to $\unit[3.6]{GHz}$. It includes an $\unit[8]{MB}$ smart cache system and a $\unit[6.4]{GT/s}$ system bus.
\item \textbf{Intel Core 2 Extreme CPU X9650}: A 64-bit, 4-core, 4-thread processor running at $\unit[3.00]{GHz}$ equipped with a $\unit[12]{MB}$ L2-cache and a $\unit[1333]{MHz}$ system bus.
\end{enumerate}

\subsection{Workload Set}
The initial workload set, mentioned in the Section \ref{fig:workloadsLoopTime}, consists of several programs from various workload suites. The programs are chosen from a variety of different application domains like scientific computing, data-mining, etc. Many of the programs are included in popular benchmarks like NAS Parallel Benchmarks, PARSEC, HPCC, NU benchmarks, etc. The programs are then executed using the $\mathsf{ELI}\text{-}\mathsf{C}$ framework to collect loop characterizations. Each program is interfaced with the largest input data set that we could find. For example, the PARSEC benchmark set provides several input sets, the largest of which is chosen. This choice ensures running the most-common execution path the largest number of times. Given the workload set, loop characterizations are collected for preprocessing. After the first data cleaning, programs with low loop-level execution are removed. Finally, programs that have restricted IP and programs whose total execution is immeasurable are removed. The remaining programs are shown in Table~\ref{tab:loopTimeChosenWorkloads}. Below is a description of the initial workload set before elimination.

The NAS Parallel Benchmarks (NPB)~\cite{barszcz1991parallel} are a well-known suite of benchmarks that proxy scientific computing applications today. A total of ten benchmarks are included in the 2.3 version release of the suite. Seven out of the ten benchmarks are used in our study. The included workloads are: CG, a conjugate gradient method used to compute an approximation to the smallest eigenvalue of a large, sparse, symmetric positive definite matrix; EP, a massively parallel kernel, generates pairs of Gaussian random deviates and tabulates the number of pairs in successive square annuli; FT, solves a 3-D partial differential equation using the fast Fourier transform (FFT); LU performs a synthetic computational fluid dynamic calculation by solving regular-sparse, block lower and upper triangular systems; MG is a multi-grid calculation; SP, similar to LU, performs a synthetic fluid dynamic problem; however, SP approaches the problem by solving multiple independent systems of non-diagonally dominant, scalar, pentadiagonal equations; finally, the UA benchmark~\cite{feng2004unstructured}, unlike the previous NAS workloads, involves some irregular and dynamically changing memory accesses as it performs computations on an unstructured adaptive mesh.

The PARSEC benchmark suite~\cite{bienia2009parsec}, developed by Princeton University, represents a diverse set of emerging applications. The PARSEC benchmarks differ from other benchmarks like SPEC in that the applications are multi-threaded in order to take advantage of multi/many-core architectures and they include emerging applications which mimic large-scale commercial programs. The set contained two out of the ten benchmarks that are included in release version 2.1, namely StreamCluster, and Freqmine. Streamcluster is a data-mining application which, given a stream of data points, finds a predetermined number of medians so that each point is assigned to its nearest center. Freqmine, also a data-mining application, employs an array-based version of the FP growth algorithm for Frequent Itemset Mining.

The HPCC benchmark version 1.4.1~\cite{luszczek2006hpc} is a set of six well known and frequently used HPC kernels: DGEMM, HPL, Random Access, FFT, Transpose and Stream. They represent a fundamental set of operations in many applications. Three workloads are selected, namely, Stream, Random Access, and Gups to test the tool's ability to identify similarity in memory access patterns. Besides workloads from these three well-known benchmark suites, independently developed parallel workloads are also included. These include two molecular dynamic workloads -- LAMMPS and MiniMD. LAMMPS~\cite{plimpton2007lammps} is an open source molecular dynamic simulator provided by the Sandia National Laboratory~\cite{sandiawebpage} and used to model atomic interaction within materials. MiniMD is part of the Mantevo application suite also provided by the Sandia National Laboratory. MiniMD is a benchmark version of LAMMPS and both use the same algorithm.

Other applications are Water N-squared simulation and K-means from SPLASH2-benchmarks~\cite{woo1995splash} provided by Stanford University. Water N-squared evaluates forces and potentials that occur over time in a system of water molecules. K-means implements the popular data mining clustering algorithm. The finite difference solver (FDS)~\cite{finitesolver} used in seismic imaging is also included, as well as a back-projection kernel used in MRI image reconstruction~\cite{takeda2010nonlinear}.

\subsection{Data Cleaning}
Starting with a total of 3000 loops, data cleaning reduces the set size down to 623 by eliminating loops having low execution times. The distribution of loops among the different workloads is shown in Table~\ref{tab:loopTimeChosenWorkloads}. This data is normalized before removing correlations. PCA, when applied, reduces the dimensions of the characterizations from 13 to 11 uncorrelated characterizations. The set is again normalized resulting in a $623\times 11$ normalized matrix on which clustering will be applied.
\begin{table*}[t]
\caption{Input workload set}\vspace{-0.1in}
\centering
\scriptsize
\label{tab:loopTimeChosenWorkloads}
\resizebox{\textwidth}{!}{\begin{tabular}{| l | l | l | c | c|}
\hline
\textbf{Workload} & \textbf{Suite} & \textbf{Description} & \textbf{\specialcell{Loop Exec.\\ Time}} & \textbf{\# Loops}\\ \hline
Water N-squared & SPLASH-2 & Evaluates the force of water molecules & 98\% & 149\\ \hline
Finite Element Solver & Internal Developer & \specialcell{Finds approximate solutions to boundary value\\ problems for partial differential equations} & 94\% & 116\\ \hline
UA & NAS & Computations on unstructured adaptive mesh & 87\% & 106\\ \hline
SP & NAS & \specialcell{Solved synthetic fluid dynamics using multiple\\ independent systems} & 93\% & 40\\ \hline
Freqmine & PARSEC & Solves FP growth algorithm & 89\% & 37\\ \hline
MRI Reconstruction & Internal Developer & MRI image reconstruction algorithm & 90\% & 24\\ \hline
CG & NAS & Computes an approximation to the smallest eigenvalue & 98\% & 24\\ \hline
MG & NAS & Multi-grid calculations& 94\% & 21\\ \hline
Lammps & \specialcell{Sandia National\\ Laboratory} & Molecular dynamic simulator & 89\% & 16\\ \hline
FT & NAS & Solves a 3-D partial differential equation using FFT & 95\% & 15\\ \hline
Streamcluster & PARSEC  & Finds nearest center to streaming points & 99\% & 15\\ \hline
Kmeans & NU-MineBench & Implements the Kmeans clustering algorithm & 89\% & 13\\ \hline
Mini MD & \specialcell{Sandia National\\ Laboratory} & \specialcell{Force computations for typical molecular\\ dynamics applications} & 96\% & 12\\ \hline
LU & NAS & Calculations of synthetic computational fluid dynamic & 98\% & 12\\ \hline
Singlestream & HPCC & Measures sustainable memory bandwidth & 97\% & 7\\ \hline
MPIRA & HPCC  & Measures the rate of integer random updates of memory & 96\% & 6\\ \hline
StarRA & HPCC & Computations on random memory accesses & 99\% & 5\\ \hline
EP & NAS & \specialcell{Generates pairs of Gaussian random deviates and\\ tabulates the number of pairs in successive square annuli} & 99\% & 5\\ \hline
\end{tabular}}
\end{table*}

\begin{figure*}[!t]
\centering
\subfloat[Workload-cluster distribution]{\includegraphics[width=2.2in]{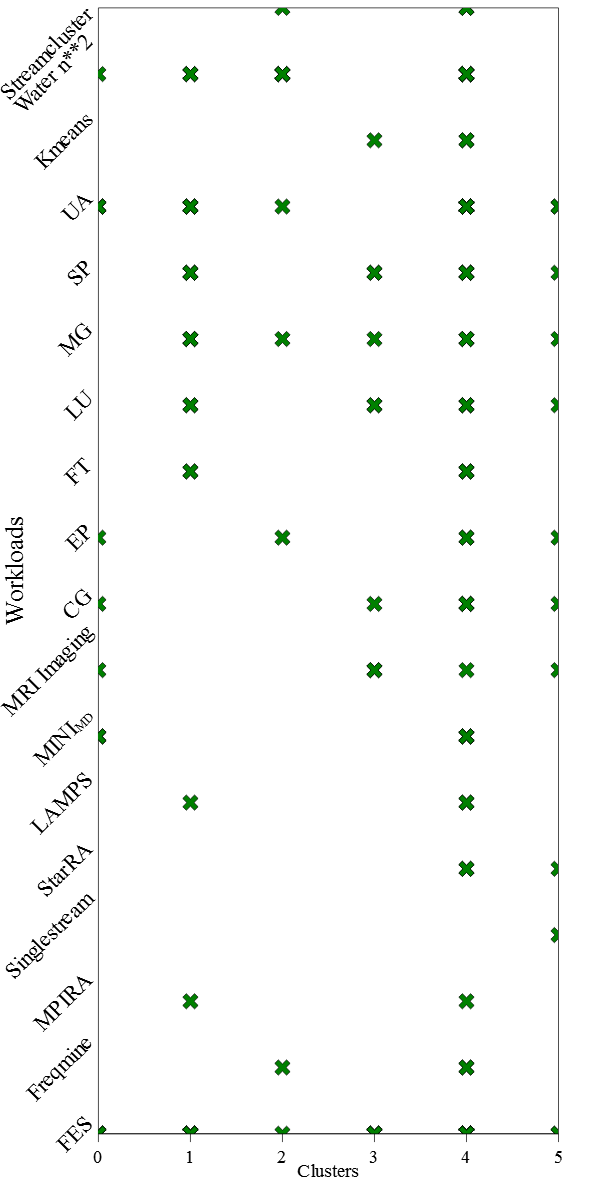}%
\label{fig:ClusterBM}}
\hfil
\subfloat[Loop distribution
among clusters]{\includegraphics[width=2.2in]{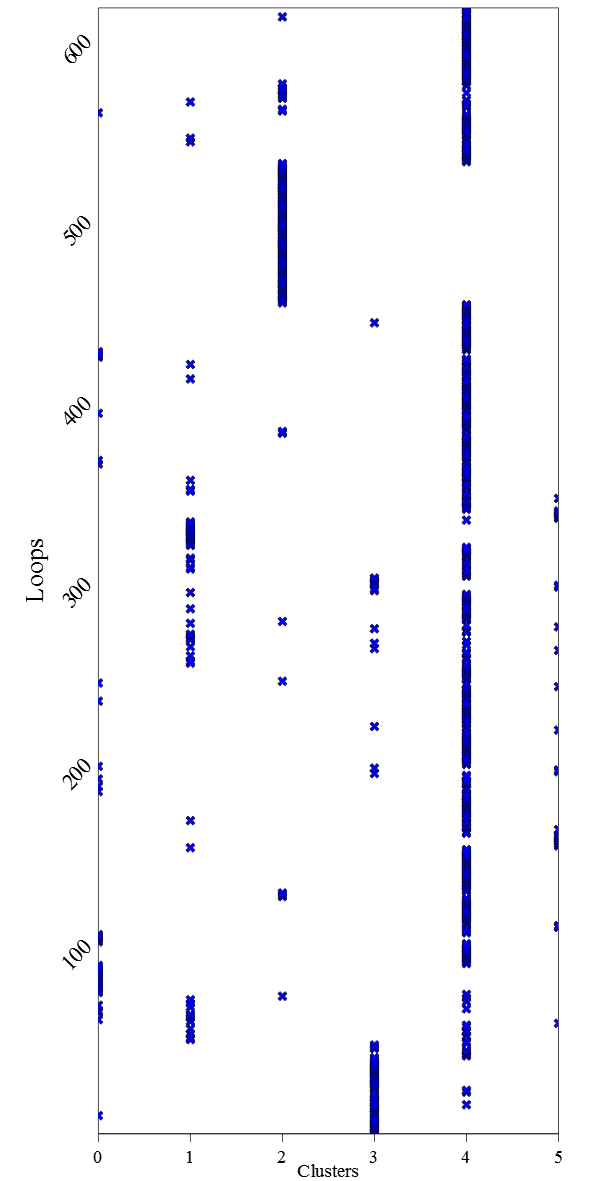}%
\label{fig:ClusterLoops}}
\caption{Workloads and loops divided into clusters}
\label{fig:clusteredData}
\end{figure*}

\subsection{Data Clustering}
Using EM algorithm from WEKA~\cite{hall2009weka}, the loops are clustered and distributed among 6 clusters. Figure~\ref{fig:ClusterBM} shows workload-cluster distribution, and Fig.~\ref{fig:ClusterLoops} shows distribution of loops among the clusters. Cluster 4 joins loops from most programs, while the remaining clusters are dispersed among the programs.
\begin{figure*}[hbtp]
\centering
\includegraphics[width=3.5in]{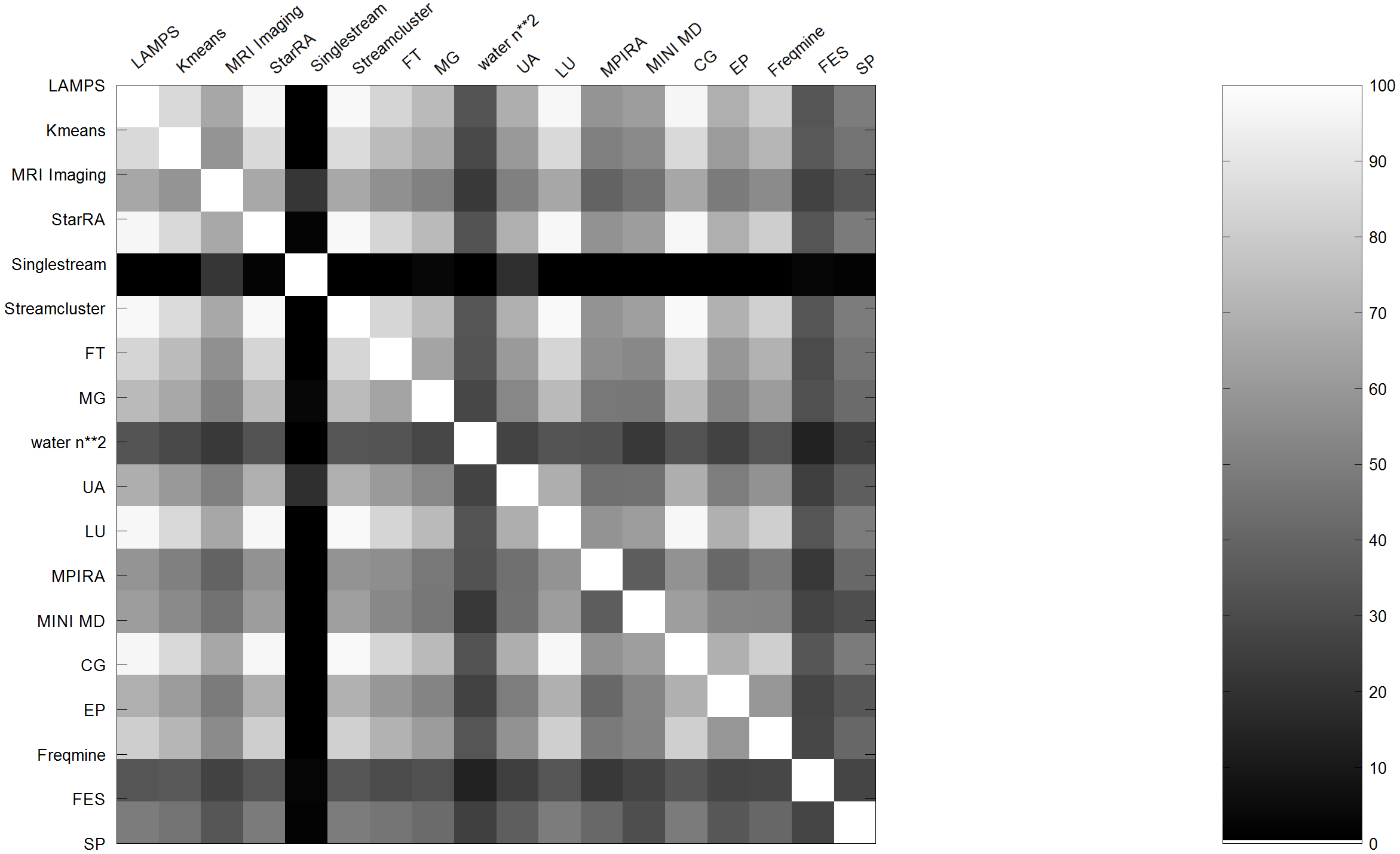}
\caption{Similarity matrix of initial workload set}
\label{fig:SimilarityMatrix}
\end{figure*}

\subsection{Calculation of Similarity Scores}
The execution time of each loop is calculated by dividing the relative latency of the executed instructions~\cite{granlund2012instruction} within the loop by the cumulative latency of the instructions executed by the entire program. The feature vector, indicating the time spent in each cluster for each program, is calculated by summing the time of all the loops in the cluster and then normalizing. For every workload-tuple, a similarity score is calculated as the inner dot product between the feature vectors of both workloads, resulting in the similarity matrix shown in Fig.~\ref{fig:SimilarityMatrix}.
\begin{figure*}[hbtp]
\centering
\includegraphics[width=5.0in]{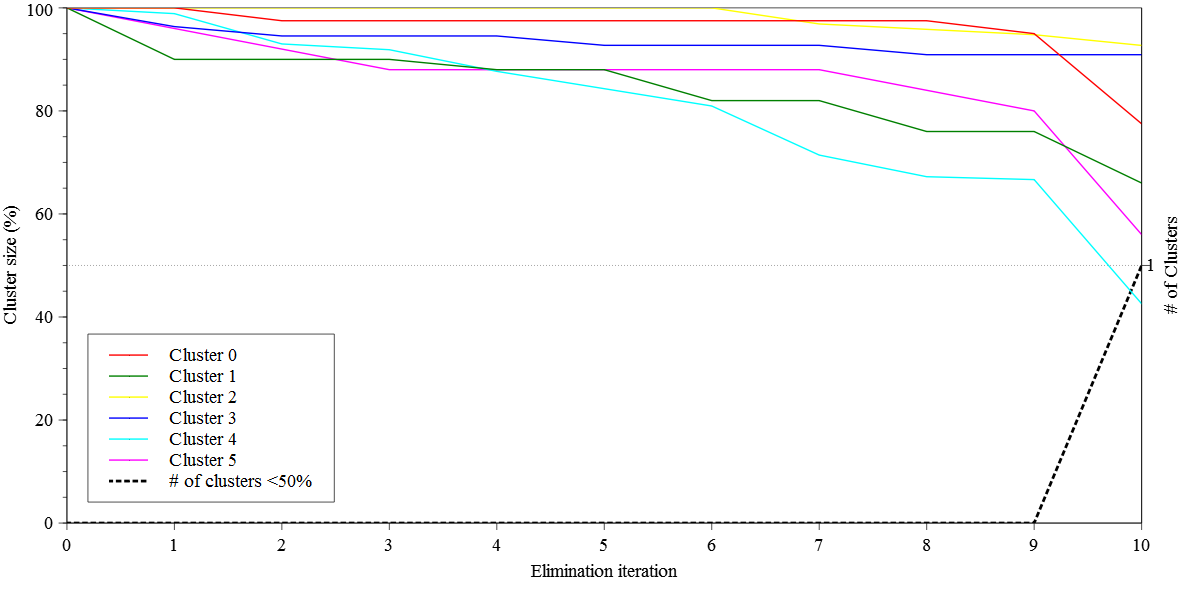}
\caption{Variation of cluster sizes after each elimination iteration}
\label{fig:EliminationStop}
\end{figure*}
\begin{figure*}[hbtp]
\centering
\includegraphics[width=3.5in]{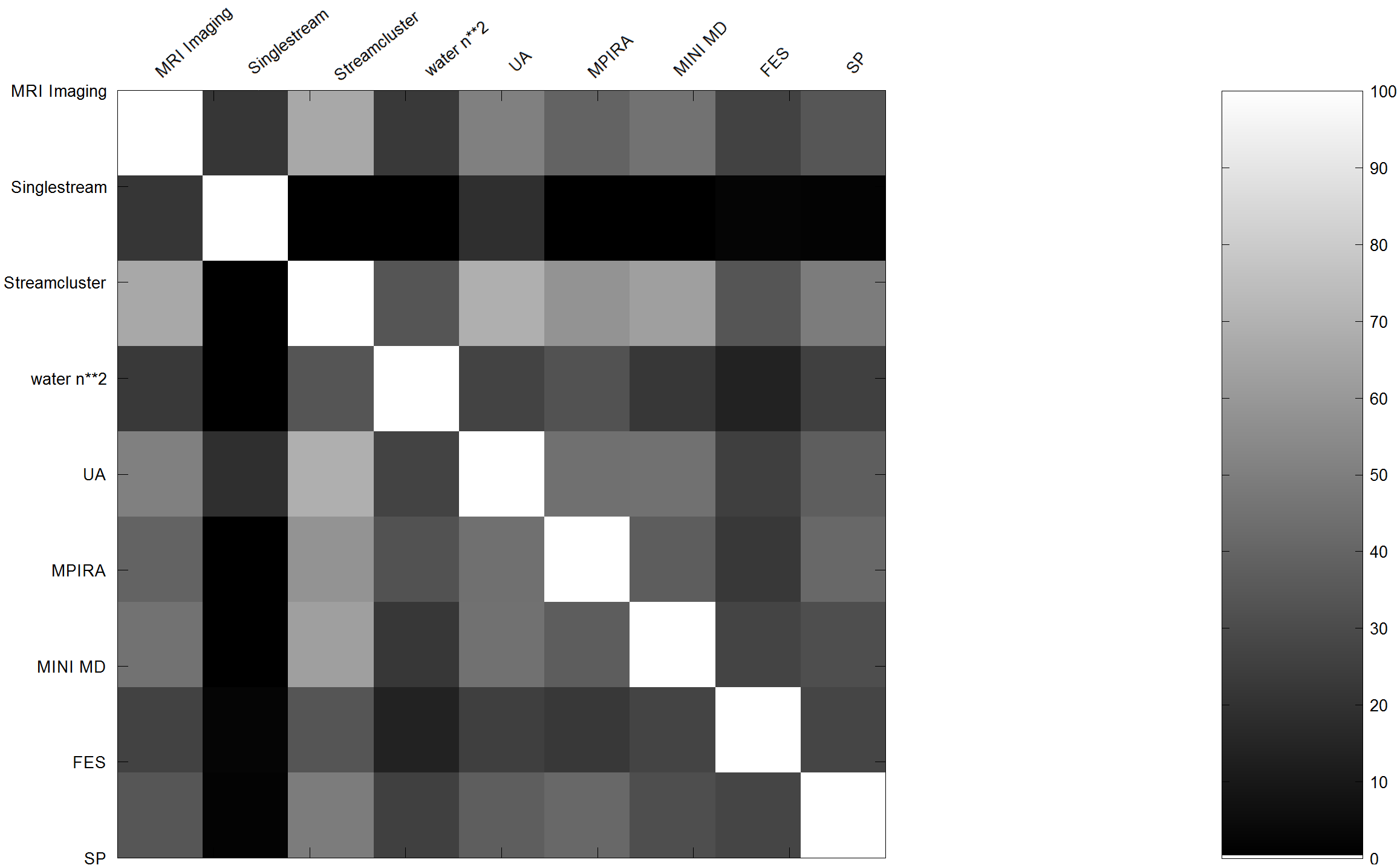}
\caption{Similarity matrix of the reduced workload set}
\label{fig:FinalSimilarityMatrix}
\end{figure*}

\subsection{Elimination}
Elimination starts from the similarity matrix. In each iteration, the workload tuple with the highest similarity score is selected. Then, the similarity of both workloads is computed against the remaining workloads. Finally, the workload with the lower similarity score is eliminated. The relative cluster size after every iteration is shown in Fig.~\ref{fig:EliminationStop}. After removing the \nth{10} workload, the size of cluster 4 is reduced to less than $50\%$ of its initial size, which is our predetermined threshold to stop elimination. So we keep the $\nth{10}$ workload and end up with a final set of size 9. The similarity matrix of the final set is shown in Fig.~\ref{fig:FinalSimilarityMatrix}.
\begin{figure*}[!t]
\centering
\subfloat[Thread speed-up]{\includegraphics[width=13pc]{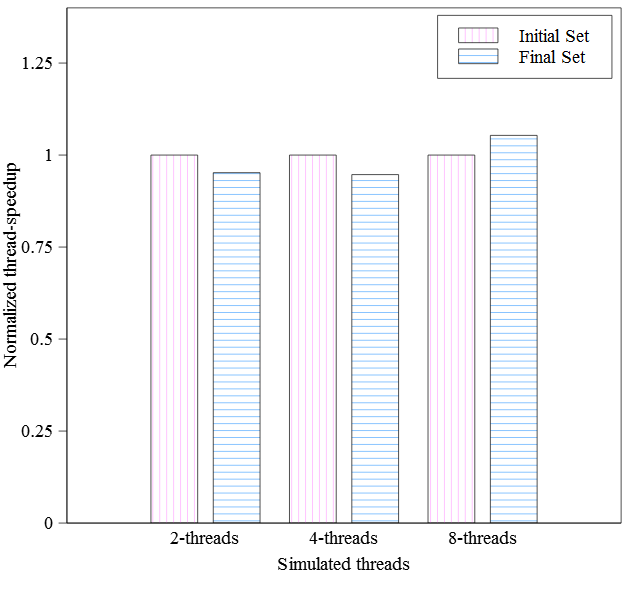}%
\label{fig:ThreadVariation}}
\hfil
\subfloat[Data sharing variation]{\includegraphics[width=13pc]{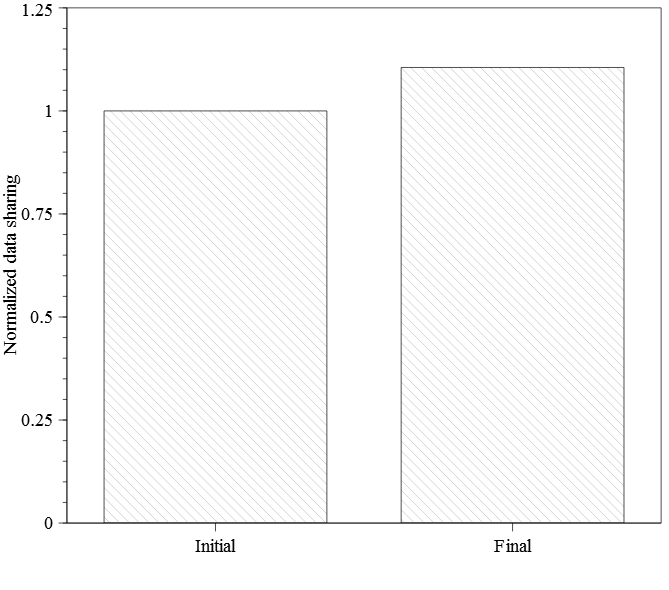}%
\label{fig:SharingVariation}}\\
\subfloat[CPI variation]{\includegraphics[width=26pc]{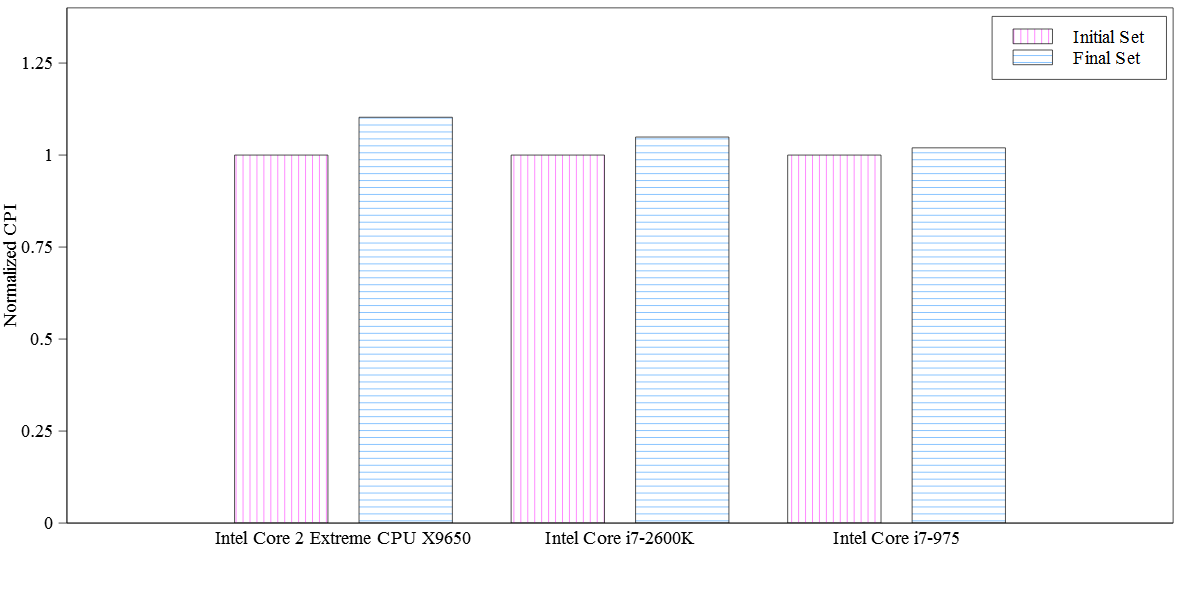}%
\label{fig:CPIVariation}}
\caption{Comparison of dynamic characteristics}
\label{fig:AllDynamic}
\end{figure*}

\begin{figure}[hbtp]
\centering
\includegraphics[width=3.3in]{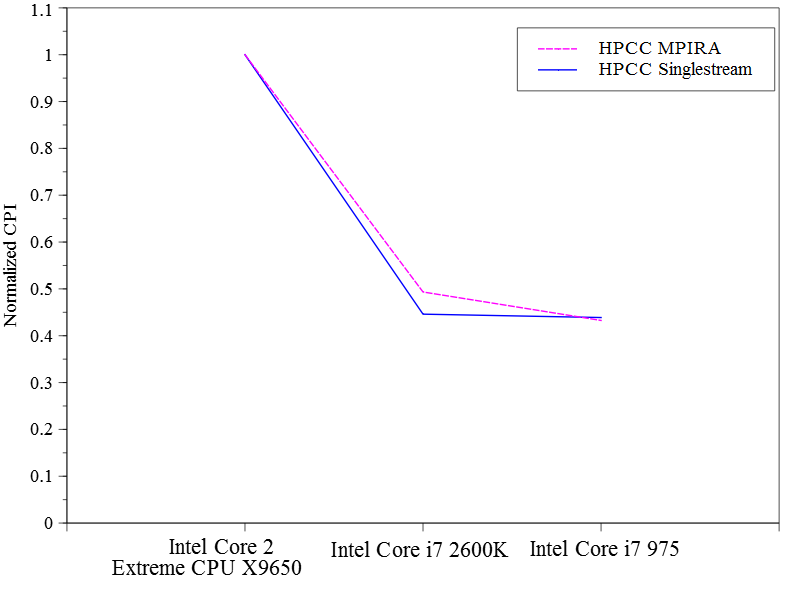}
\caption{HPCC bottleneck on older Intel generations}
\label{fig:CPIPenryn}
\end{figure}

\begin{figure}[!t]
\centering
\subfloat[Intel Core i7-975]{\includegraphics[width=16pc]{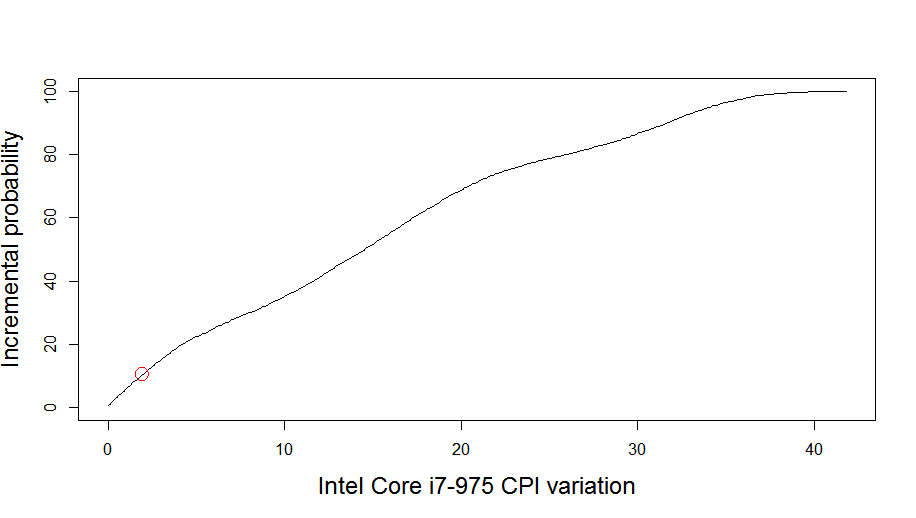}%
\label{fig:random0}}\\
\subfloat[Intel Core i7-2600K]{\includegraphics[width=16pc]{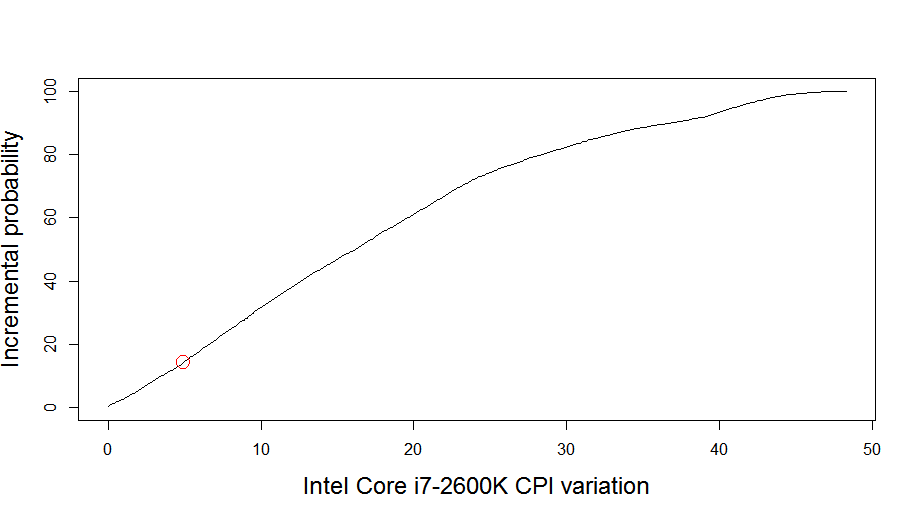}%
\label{fig:random1}}\\
\subfloat[Intel Core 2 Extreme CPU X9650]{\includegraphics[width=16pc]{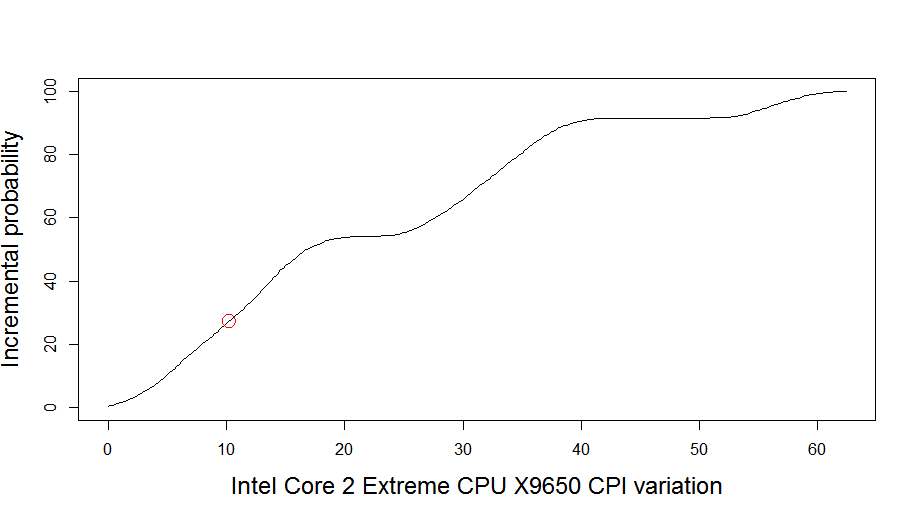}%
\label{fig:random2}}
\caption{CPI variation from random 9 workload sets}
\label{fig:randomAll}
\end{figure}

\subsection{Validation}
The elimination scheme is able to reduce the set by 50\% of its original size. Cluster 4 is the dominant cluster before elimination, which justifies its quick reduction. To verify that the reduced set is representative of the initial set, the programs are executed on the aforementioned Intel processors. The following characterizations are collected: CPI, data sharing, and total execution cycles when run with 1, 2, 4, and 8 threads. The characteristics of the final set are then compared to those of the initial set, as shown in Fig.~\ref{fig:AllDynamic}. Numerically, the CPI varies between 10\%, 4\%, 1\% on Intel Core 2 Extreme CPU X9650, Intel Core i7-2600K, and Intel Core i7-975, respectively as shown in Fig.~\ref{fig:CPIVariation}. The highest difference is spotted on the Intel Core 2 Extreme CPU X9650. After studying the behavior of the programs on this system, the difference is attributed to the fact that Intel Core 2 Extreme CPU X9650 is the oldest technology among processors tested, and some of the programs showed slow execution. This can be observed by comparing the CPI of HPCC MPIRA and HPCC SingleStream on the three generations of processors. As shown in Fig.~\ref{fig:CPIPenryn}, the CPI is reduced by more than half between both generations. This CPI drop is the highest drop between programs in the set. Since the workloads are simulated on the higher performance processor and there is a CPI drop between simulated processors, then results collected on the lower performing processor are expected to differ.  The thread speed-up on Intel Core i7-2600K varied between 4\% and 5\%, as shown in Fig.~\ref{fig:ThreadVariation}. Finally, the variation in data sharing is around 10\%, as shown in Fig.~\ref{fig:SharingVariation}. The final subset attains a highly similar CPI and thread speedup scores, while a larger margin of error is achieved by data sharing. This can be justified by the nature of the collected characterizations, which lacked any value to represent data sharing.

To validate that our methodology is superior to random picking, a lengthy simulation with 10,000 iterations is run. At each iteration, 9 programs are randomly picked and their CPI variation from the initial set is calculated. As marked using a round red dot in Fig.~\ref{fig:randomAll}, there is a 10\% probability of getting a CPI variation similar or better than ours in both Fig.~\ref{fig:random0} and Fig.~\ref{fig:random1} on Intel Core i7-975 and i7-2600K. The biggest difference of around 25\% in Intel Core 2 Extreme shown in Fig.~\ref{fig:random2} is also attributed to the difference between generations of processors as discussed previously. 

\section{Micro-benchmarks}
Upon validating our initial hypothesis we conclude that characterizing a single iteration of loop-centric workloads is sufficient to understand the behavior of workloads. Hence, a loop-dominant workload set can be represented by the loops within the different workloads. In this section, we check the feasibility of isolating the loops from a program and automatically generating a new micro-workload having the same instruction mix as the loops in the initial program.

Compute-centric applications are one type of loop-dominant workloads since they rely on manipulating large amounts of data by performing certain computations on them. These computations can be viewed from two different perspectives. Firstly, from a program's perspective, the accuracy of the final output is crucial to verify the program's correctness regardless of the type of instructions running the computation. For example, given inputs $A$ and $B$, the execution of an addition function should always generates the output $(A+B)$, regardless of the type and number of instructions used. This matter is crucial for software testing.

Secondly, from a processor's perspective, the computations performed (namely executed instructions and their data sources such as registers and memory) are more important than knowing the actual values of $A$ and $B$ and testing the accuracy of their sum. Therefore, the correctness of the result is not of major concern to the processor since it is interpreted from a programmer's point of view. As a result, collecting the set of executed assembly instructions and ensuring that they execute would guarantee, from a processor's point of view, that the behavior of the original workload is faithfully emulated.

In order to achieve this task, a tool is developed and published online (\url{https://bitbucket.org/shacs/ubenchmark}). Given any x86 binary compiled with debugging (-pg) flags, this tool generates a new C++ micro-workload having identical loop instruction execution mix as the initial program. The input data set will be random, but as previously discussed, this issue is not a major concern from a characterization perspective.

\subsection{Work flow}
Similar to the previously described tool $\mathsf{ELI}\text{-}\mathsf{C}$, binaries given to the tool script are simulated using Valgrind to generate the assembly line execution counts. Each loop is then isolated and mapped to a standalone function in the micro-workload. Control instructions in the initial program in Fig.~\ref{fig:AutomatedFoorLoopInput} are transformed as shown Fig.~\ref{fig:AutomatedFoorLoopOutput}.

The generated micro-workload contains $n$ functions, where $n$ is the total number of standalone loops in the initial program. Within each function, a loop iterates over the number of iterations executing the loop's initial assembly instructions.

\begin{figure}[t]
	\centering
	\begin{tabular}{c}
\begin{lstlisting}[lastline=1]
for(temp=0;temp<2727;temp++)
\end{lstlisting}
	\end{tabular}
\caption{Initial program \textit{for loop} condition}
\label{fig:AutomatedFoorLoopInput}
\end{figure}

\begin{figure}[t]
	\centering
	\begin{tabular}{c}
\begin{lstlisting}[lastline=8]
\\Start of loop
__asm__("mov \$0, %ecx;\n.L0\n")

\\End and increment loop
__asm__("addl   \$0x1, %ecx;”)
__asm__("cmp    \$AA7, %ecx;")
__asm__("jle    .L0;")
\end{lstlisting}
	\end{tabular}
\caption{Automated micro-benchmark \textit{for loop} condition}
\label{fig:AutomatedFoorLoopOutput}
\end{figure}

\subsection{Avoiding Pitfalls}
Compilers perform complex operations when transforming source code to binary. It is an oversimplification to simply detach excerpts of assembly code and combine them in a new workload and expect it to work. Since compilers are platform dependent, the generated C++ file should be processed when migrating to a new platform. The following section introduces scenarios that should be avoided to ensure reproducing an executable micro-benchmark.

\subsubsection{Loop bounds}
As shown in Fig.~\ref{fig:AutomatedFoorLoopOutput}, the automatically generated loop bounds depend on \%ecx register. As a result, if any instruction in the loop body alters the \%ecx register it modifies the loop's execution count. Since we only care about hardware correctness, then replacing \%ecx register with another register suffices to overcome this issue.

\subsubsection{Nested Loops/Conditionals within Loops}
Although nested loops and conditionals within loops might seem uneasy to imitate, yet as previously discussed, breaking non-linear loops into code blocks that execute linearly depending on the number of times each path executes solves the issue. This approach preserves the number of times each assembly line is executed. However, it is not an easy job to do when these nested structures execute the same number of times. This results in a single loop body with a conditional jump in it. Since data is randomly accessed in the micro-workload, there is no guarantee that it executes the same number of times. In this case processing the body of the loop and removing all known jump instructions ensures execution correctness.

\subsubsection{Out-of-Range Memory Accesses}
Memory accesses present another problem. Both writes and reads should access the same location they were intended to access in order to preserve the original program's memory usage patterns. However, since memory is dependent on the overhead instructions that the compiler generates, it is difficult to reproduce the behavior exactly, and this might lead to segmentation faults. In this case, preprocessing all memory access and ensuring the offset is valid, is a required step.

\begin{table*}[t]
\caption{Workload versus micro-workload}
\centering
\label{tab:Workloads_vs_microworkloads}
\scriptsize
\resizebox{\textwidth}{!}{\begin{tabular}{| c | c | c | c | c|}
\hline
\textbf{\specialcell{Workload}} & \textbf{No. of Loops} & \textbf{Workload Instruction Count}& \textbf{Micro-workload Instruction Count}\\ \hline\hline
Freqmine & 57&500,417,113,277 & 35,457,906 \\ \hline
Canneal& 17& 53,318,055,570& 4,134,073 \\ \hline
Streamcluster& 11&139,424,798,411 & 3,558,162 \\ \hline
Swaptions& 15& 37,855,234,863& 1,319,935 \\ \hline
Blackscholes& 4& 7,776,110,497& 2,712,197 \\ \hline
\end{tabular}}
\end{table*}

\subsection{Simulation Results}
To validate the effectiveness of our micro-workload generator, we simulate 5 programs from the PARSEC benchmark suite, and after post-processing with a platform script, we obtain 5 executable micro-workloads. The number of loops in the program as well as a comparison between the program and micro-workload is shown in
Table~\ref{tab:Workloads_vs_microworkloads}.

Upon inspection, the micro-workloads match the initial programs' loop execution counts and instruction mix. The transformation between workloads and micro-workloads is beneficial to save simulation time when investigating a benchmark's instruction mix. It is also feasible to join multiple micro-benchmarks in the same program thus reducing execution of multiple benchmarks into the same benchmark.

\section{Conclusions}
A novel methodology that reduces the size of a workload set used to characterize a processor's performance has been proposed. The methodology, which is targeted for loop-dominant programs, characterizes programs in the workload set, and then eliminates similar programs. First, characterizations are used to form a feature vector that provides a signature for each loop. By processing all such vectors through machine learning techniques, loops are categorized into categories having similar loop structures. Remapping workloads into a unique set of loop structures allows generating a signature vector for each program. Similarity between program pairs is then calculated as the inner product of the signature vectors of both programs. Using a systematic elimination process on the similarity scores, an initial set of workloads can then be reduced to a smaller representative subset. This approach reduces simulation time solely based on the fact that loops are the essential driving engine for most compute intensive workloads, and thus they should be driving the subsetting algorithm.

While the proposed framework reduces the size of a workload set in order to reduce simulation time, this framework has been also extended to generate micro-benchmarks that are unique and representative of any workload set. These micro-benchmarks can be used in simulations for processor design as well. Overall, by characterizing workloads at the loop-level, the proposed framework not only achieves more accurate and meaningful subsetting of the workloads, but also opens up the possibility to leverage distinct loops for exploring the behavior of emerging workloads.

\bibliographystyle{IEEEtran}
\bibliography{IEEEabrv,MyReferences}

\begin{IEEEbiography}[{\includegraphics[width=1in,height=1.25in,clip,keepaspectratio]{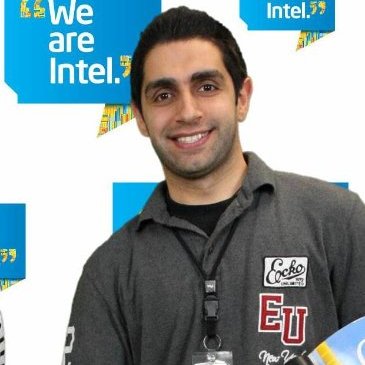}}]{Elie Shaccour} received the B.E. degree in computer and communications engineering (CCE) from the American University of Beirut (AUB), Beirut, Lebanon, in 2014, and his Master's degree in CCE from AUB in 2014. He is currently working as a programmer analyst at Audi, Lebanon. His current research interests include computer architecture, workload characterization, software management, and software testing.
\end{IEEEbiography}
\vfill
\begin{IEEEbiography}[{\includegraphics[width=1in,height=1.25in,clip,keepaspectratio]{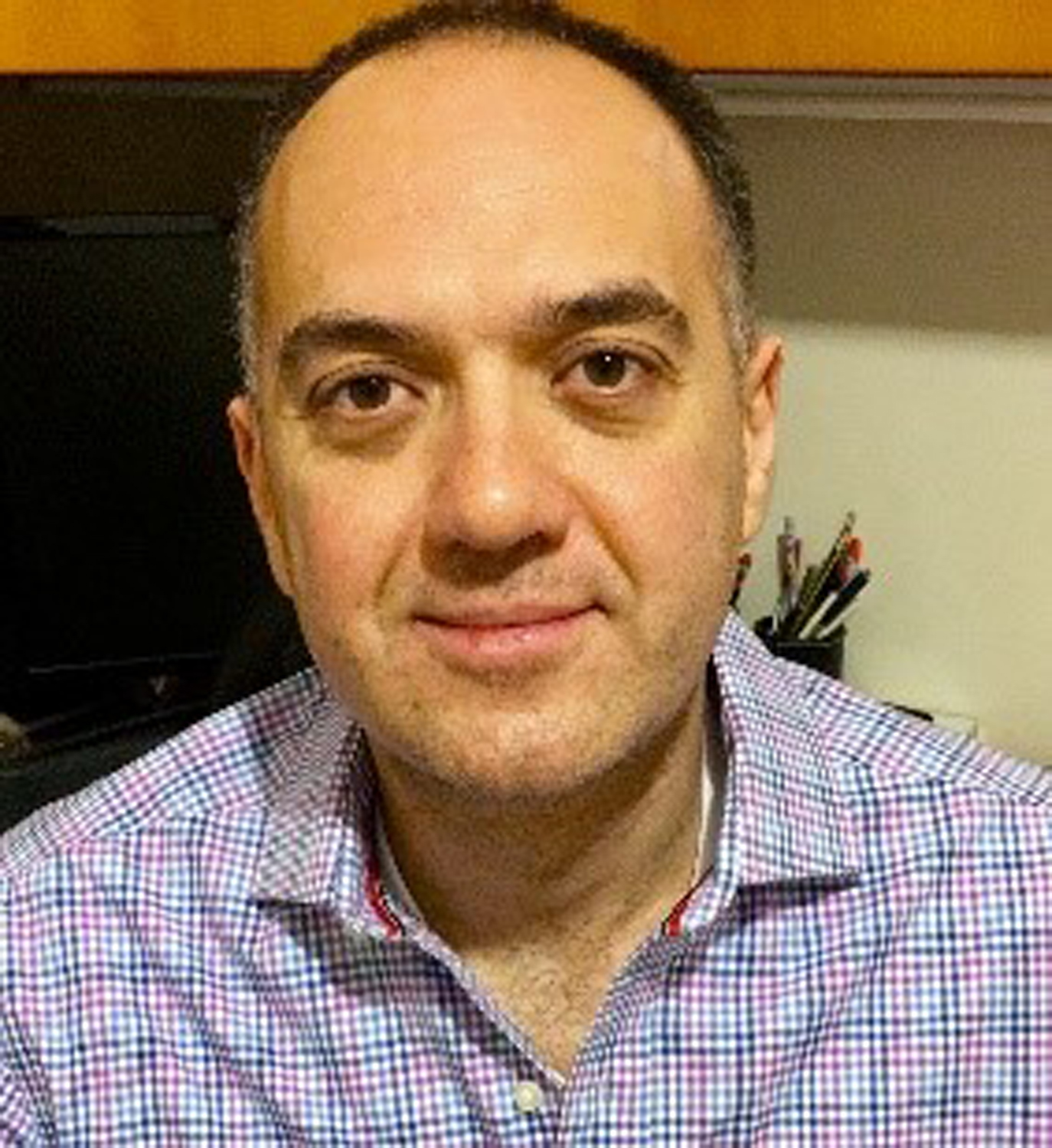}}]{Mohammad M. Mansour} (S'97-M'03-SM'08) received the B.E. (Hons.) and the M.E. degrees in computer and communications engineering from the American University of Beirut (AUB), Beirut, Lebanon, in 1996 and 1998, respectively, and the M.S. degree in mathematics and the Ph.D. degree in electrical engineering from the University of Illinois at Urbana–Champaign (UIUC), Champaign, IL, USA, in 2002 and 2003, respectively.

\noindent He was a Visiting Researcher at Broadcom, Sunnyvale, CA, USA, from 2012 to 2014, where he worked on the physical layer SoC architecture and algorithm development for LTE-Advanced. He was on research leave with Qualcomm Flarion Technologies in Bridgewater, NJ, USA, from 2006 to 2008, where he worked on modem design and implementation for 3GPP-LTE, 3GPP2-UMB, and peer-to-peer wireless networking physical layer SoC architecture and algorithm development. He was a Research Assistant at the Coordinated Science Laboratory (CSL), UIUC, from 1998 to 2003. He worked at National Semiconductor Corporation, San Francisco, CA, with the Wireless Research group in 2000. He was a Research Assistant with the Department of Electrical and Computer Engineering, AUB, in 1997, and a Teaching Assistant in 1996. He joined as a faculty member
with the Department of Electrical and Computer Engineering, AUB, in 2003, where he is currently an Associate Professor. His research interests are in the area of energy-efficient and high-performance VLSI circuits, architectures, algorithms, and systems for computing, communications, and signal processing.

\noindent Prof. Mansour is a member of the Design and Implementation of Signal Processing Systems (DISPS) Technical Committee Advisory Board of the IEEE Signal Processing Society. He served as a member of the DISPS Technical Committee from 2006 to 2013. He served as an Associate Editor for {\sc IEEE Transactions on Circuits and Systems II} (TCAS-II) from 2008 to 2013, Associate Editor for the {\sc IEEE Transactions on VLSI Systems} from 2011 to 2016, and Associate Editor for the {\sc IEEE Signal Processing Letters} from 2012 to 2016. He served as the Technical Co-Chair of the IEEE Workshop on Signal Processing Systems in 2011, and as a member of the Technical Program Committee of various international conferences and workshops. He was the recipient of the PHI Kappa PHI Honor Society Award twice in 2000 and 2001, and the recipient of the Hewlett Foundation Fellowship Award in 2006. He has seven issued U.S. patents. For further information on research in progress and associated publications please refer to \url{http://staff.aub.edu.lb/~mm14/}.
\end{IEEEbiography}

\end{document}